\documentclass[fleqn,10pt,twocolumn]{wlscirep} %
\usepackage[utf8]{inputenc}
\usepackage[T1]{fontenc}

\usepackage{subfigure} 
\usepackage[abs]{overpic}
\usepackage{chemformula} 

\usepackage{placeins} 
\usepackage{capt-of} 

\newcommand{\beginsupplement}{%
	\setcounter{table}{0}
	\renewcommand{\thetable}{S\arabic{table}}%
	\setcounter{figure}{0}
	\renewcommand{\thefigure}{S\arabic{figure}}%
}

\title{Isotope Exchange Raman Spectroscopy (IERS): a novel technique to probe physicochemical processes \textit{in situ}}
\author[1,*]{Alexander Stangl}
\author[1]{Dolors Pla}
\author[2]{Caroline Pirovano}
\author[1]{Odette Chaix-Pluchery}
\author[3,4]{Federico Baiutti}
\author[3]{Francesco Chiabrera}
\author[3,5]{Albert Tarancón}
\author[1]{Carmen Jiménez}
\author[6]{Michel Mermoux}
\author[1,*]{Mónica Burriel}
\affil[1]{Univ. Grenoble Alpes, CNRS, Grenoble-INP, LMGP, 38000 Grenoble France}
\affil[2]{Univ. Lille, CNRS, Centrale Lille, Univ. Artois, UMR 8181 – UCCS – Unité de Catalyse et Chimie du Solide, F-59000 Lille, France}
\affil[3]{Catalonia Institute for Energy Research (IREC), Barcelona, Spain}
\affil[4]{Departement of Materials Chemistry, National Institute of Chemistry, Hajdrihova 19, Ljubljana SI-1000, Slovenia}
\affil[5]{ICREA, 23 Passeig Lluis Companys, 08010 Barcelona, Spain}
\affil[6]{Univ. Grenoble Alpes, Univ. Savoie Mont Blanc, CNRS, Grenoble INP, LEPMI, 38000, Grenoble, France }
\affil[*]{alexander.stangl@grenoble-inp.fr, monica.burriel@grenoble-inp.fr}

\begin{abstract}
We have developed a novel \textit{in situ} methodology for the direct study of mass transport properties in oxides with spatial and unprecedented time resolution, based on Raman spectroscopy coupled to isothermal isotope exchanges. 
Changes in the isotope concentration, resulting in a Raman frequency shift, can be followed in real time, not accessible by conventional methods, enabling complementary insights for the study of ion transport properties of electrode and electrolyte materials for advanced solid-state electrochemical devices. 

The proof of concept and strengths of isotope exchange Raman spectroscopy (IERS) are demonstrated by studying the oxygen isotope back-exchange in gadolinium-doped ceria (CGO) thin films. 
Resulting oxygen self-diffusion and surface exchange coefficients are compared to conventional time-of-flight secondary ion mass spectrometry (ToF-SIMS) characterisation and literature values, showing good agreement, while at the same time providing additional insight, challenging established assumptions.
IERS captivates through its rapidity, simple setup, non-destructive nature, cost effectiveness and versatile fields of application and thus can readily be integrated as new standard tool for \textit{in situ} and \textit{operando} characterization in many laboratories worldwide.
The applicability of this method is expected to consolidate our understanding of elementary physicochemical processes and impact various emerging fields including solid oxide cells, battery research and beyond.

\end{abstract}

\begin{document}

\flushbottom
\maketitle

\thispagestyle{empty}
\section*{Introduction}
Many types of advanced solid state electrochemical devices, spanning over a large range of applications from energy conversion, transport and storage, environmental gas sensing to neuromorphic computing, are built upon versatile ionic and/or electronic conducting materials, where ion kinetics are key for the functional properties. Oxides are at the center of this emerging materials class \cite{Sun2021}. 
Accelerating the development of tuned metal oxides is therefore an integral part towards high performance and miniaturized electrochemical systems. 
This strategy requires innovative approaches of materials synthesis \cite{Acosta2019} and nano-engineering \cite{Stangl2021d}, as well as a constantly expanding toolbox for deeper fundamental understandings. 
Time and cost intensive techniques based on large scale facilities or lab based cutting edge experimental setups (\textit{e.g.} synchrotron, \textit{in situ} TEM \cite{Hwang2019}, NAP-XPS, \textit{etc.} \cite{Opitz2017b}) contribute with outstanding insights into highly relevant physicochemical processes with remarkable spatial and temporal resolution. However, the widespread availability of advanced characterisation techniques, especially focusing on \textit{in situ} and \textit{operando} characterisation \cite{Li2018c,Stangl2021b,Meyer2019}, is essential for the efficient and timely material studies with high throughput capabilities.  
Traditionally, oxygen transport properties are addressed by three approaches: electrochemical impedance spectroscopy (EIS), electrical conductivity relaxation (ECR) and isotope exchange depth profiling coupled to secondary ion mass spectrometry (IEDP-SIMS). Each one is characterised by certain benefits and limitations in terms of applicability, costs, information depth and preservation of sample integrity.
Here, we aim to expand the scientific toolkit by introducing a novel, readily accessible \textit{in situ} methodology based on the coupling of isotopic exchange with micro Raman spectroscopy for the characterization of mass transport kinetics in real time, without the need of expensive and/or time-consuming techniques.

Raman spectroscopy is known as a powerful and efficient, yet standard tool for assessing the microstructure and defect landscape of various types of materials including metal oxides. 
In particular it has a long lasting and successful history for \textit{in situ} characterization of catalyst materials \cite{oyama1999,Kuba2002,Mestl2013} and, more recently, for lithium intercalation in battery electrodes \cite{Migge2005,Stancovski2014} using visible light.
Raman spectroscopy is generally considered as a vibrational spectroscopy technique and, as such, sensitive to the mass of the atoms constituting a crystal. 
The isotopic substitution thus provides a mass contrast, which can be utilized for the assignment of vibrational modes \cite{Kim1997}, the study of isotopic fractions \cite{Guerain2015} or irradiation effects \cite{Ciszak2019} and to investigate bevelled surfaces for isotopic depth profiling \cite{Kim1994,Stender2013}. 
However, there are only a few works that combine \textit{in situ} Raman spectroscopy with $^{18}$O/$^{16}$O isotopic substitution \cite{Mestl1994,Mestl1994a}, focussing on surface structures \cite{Tsilomelekis2010,Lee2008,Weckhuysen2000}, rather than exchange kinetics. Oxygen self-diffusion in Y$_2$O$_3$-ZrO$_2$ bulk has only been addressed indirectly by Raman spectroscopy by following the amount of $^{18}$O tracer atoms in the ambient gas \cite{Kim1993}.
Finally, and only very recently, \textit{in situ} Raman measurements have been employed to extract kinetic coefficients via hydrogen/deuterium exchange by tracking the fraction of Raman band intensities \cite{MoreiraDaSilva1997,Gao2020,Gao2021a}.

In this work, we take advantage of the mass sensitivity of Raman spectroscopy and demonstrate the possibility to follow continuous temporal changes in the vibrational frequency spectra as a function of the isotopic concentration of oxygen by \textit{in situ} Raman spectroscopy for the characterisation of tracer exchange and diffusion properties and thus, make oxygen movement inside a material visible. 
To the best of our knowledge, this is the first time that oxygen tracer exchange processes can be directly followed in real time by any means.
This methodology can be readily expanded to other fields of application, including \textit{e.g.} the study of isotopically labelled Li in Li-ion batteries and O in new O-ion batteries \cite{Schmid2023}, heralding a new era of tracer-based experiments.


In the following section, the fundamentals of the isotope effect on the Raman spectra are briefly introduced and the novel \textit{in situ} isotope exchange Raman spectroscopy (IERS) methodology is presented, including a discussion on requirements, advantages and limitations.
Finally, surface exchange coefficients with corresponding activation energies and in-plane diffusion coefficients for Ce$_{0.8}$Gd$_{0.2}$O$_{2-\delta}$ thin films are reported for \textit{in situ} measurements and compared to values obtained by the conventional ToF-SIMS approach and literature. 

\begin{figure}[t]
	\centering
	\includegraphics[width=80mm, unit=1mm]{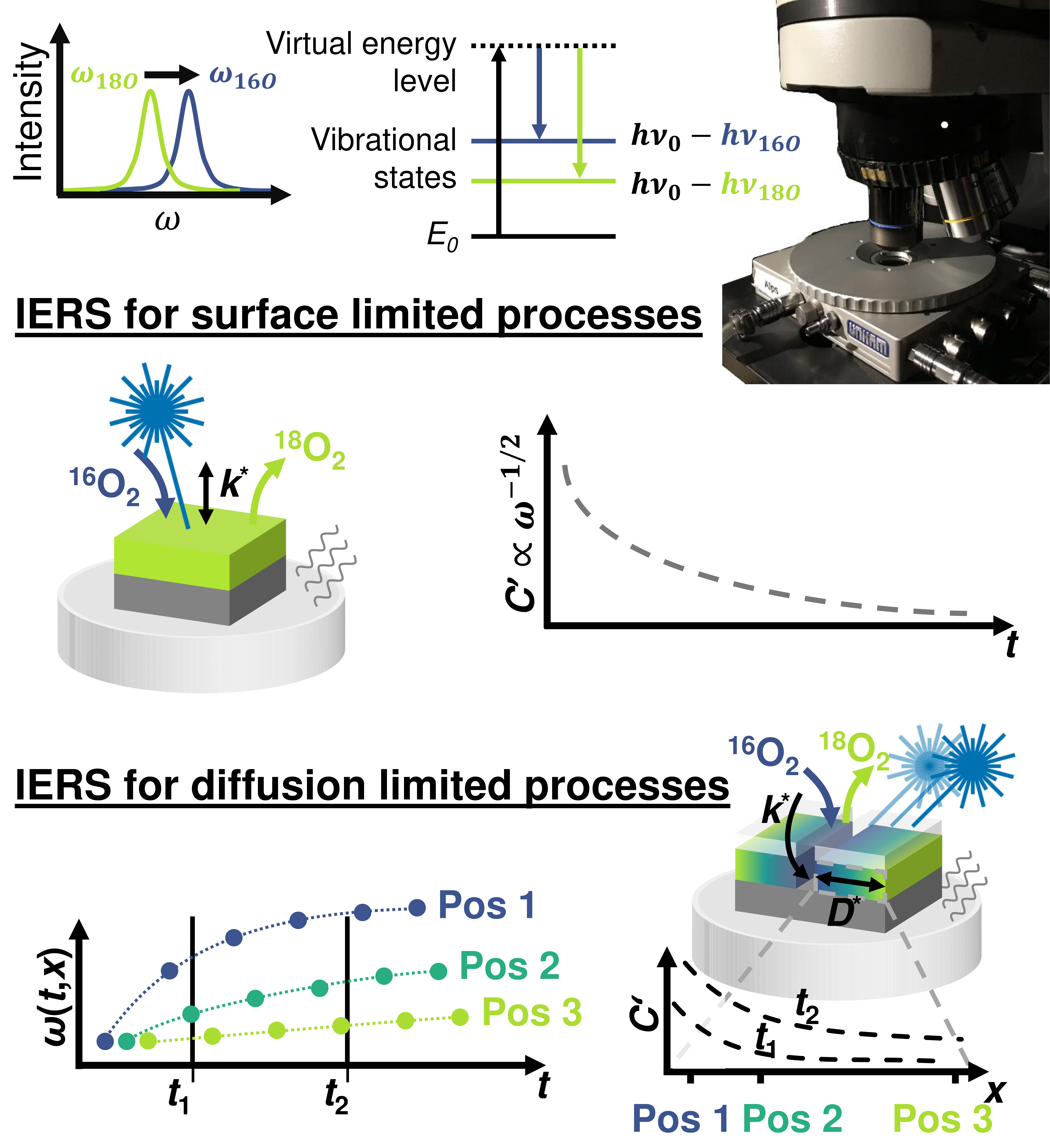}
	\caption[]{Schematic of novel IERS methodology to study surface and diffusion limited processes, based on isotopic Raman shifts.}
	\label{fig:IERS_schematic}
\end{figure}

\section*{Methods}
\subsection*{Isotope effect in Raman spectroscopy}

As vibrational spectroscopy technique Raman spectroscopy is sensitive to the mass of the atoms constituting a molecule or a crystal. Within the framework of the harmonic approximation, the wavenumber, $\omega$, of a specific vibrational mode is proportional to:
\begin{equation}\label{eqn:harmonic_wavenumber}
	\omega \propto \sqrt{\frac{k}{\mu}},
\end{equation}
where $k$ is a spring constant and $\mu$ the reduced mass. Here, the effective reduced mass provides chemical and/or isotopic information. The virtual crystal approximation (VCA) is usually introduced for a crystal containing several isotopes in order to recover the translational invariance lifted by the isotopic disorder, meaning that masses of these isotopes are simply replaced by their average weighted relative abundances, $\bar{m}$ \cite{Cardona2005,Mermoux2021}.
This approximation allows to understand the \textit{one mode behavior} observed for monoatomic crystals as well as for crystals with different elements in the primitive cell, \textit{i.e.} the isotopic mass disorder is only a weak perturbation of the crystal properties. 

While IERS can be applied using different isotopic tracer elements, including $^{6}$Li and $^{18}$O, we will restrict ourselves in the following to the latter.
The analysis for O-dominated or purely oxygen related vibrational modes is straightforward. The mass of the virtual crystal atoms is given by:
\begin{equation}\label{eqn:Intro:VCA_mass}
	\bar{m}={{^{18}C}}\cdot m_{^{18}\text{O}} + (1-{{^{18}C}})m_{^{16}\text{O}},
\end{equation}
with the atomic weights $m_{^{18}\text{O}}=17.999$\,u and $m_{^{16}\text{O}}=15.995$\,u of $^{18}$O and $^{16}$O, respectively and the $^{18}$O isotopic fraction: 
\begin{equation}\label{eqn:Intro:OIF}
	{^{18}C}=\frac{[^{18}\text{O}]}{[^{18}\text{O}]+[^{16}\text{O}]}.
\end{equation}


Since the force constants, in the harmonic approximation, are constant, the relative shift due to isotopic substitution, can be estimated within the VCA model using Eq.~\ref{eqn:harmonic_wavenumber}:
\begin{equation}\label{eqn:Intro:VCA_fit}
	\frac{\omega ({^{18}C})}{\omega_\text{ref}} = \sqrt{\frac{\mu_\text{ref}}{\mu({^{18}C})}} = \sqrt{\frac{\bar{m}_\text{ref}}{\bar{m}({^{18}C})}},
\end{equation}
and thus  $\omega \propto {^{18}C}^{-1/2}$.


\subsection*{\textit{In situ} isotope exchange Raman spectroscopy}
IERS is based on the sensitivity of Raman spectroscopy to isotopic fractions.
Upon thermal activation of oxygen isotopic exchange by heating a sample in a gas atmosphere of isotopic concentration different from the specimen, the Raman shift can be followed \textit{in situ} with time and spatial resolution (see Fig.~\ref{fig:IERS_schematic}).
IERS can be applied to study the exchange (\textit{i.e.} introducing $^{18}$O into the sample) as well as the back-exchange (\textit{i.e.} removing $^{18}$O from the sample) as a function of temperature, pressure, humidity, gas constituents, \textit{etc} in a non-destructive manner. Studying the back-exchange, however, allows to drastically reduce the tracer gas consumption and enables the use of conventional temperature cells. To obtain meaningful transients, exchanges must be performed under isothermal conditions. 

We developed two scenarios for the application of \textit{in situ} IERS. The first one, for the direct characterization of the tracer surface exchange coefficient, $k^*$, is restricted to the study of surface limited reactions and thus to thin films and (nano-) powders. The measured transients can be related to a kinetic model, as outlined in the next section. 

The second case expands the applicability of IERS to diffusion limited reactions (thin films and dense bulk materials) and gives access as well to the in-plane diffusion coefficient, $D^*$. Therefore the sample is coated with a thin, Raman transparent and inert conformal capping layer to inhibit oxygen exchange at the surface.
Next, a trench in the coating and film/bulk is opened, exposing the materials surface to the atmosphere and thus allowing oxygen exchange only along one well defined side \cite{footnote1}.
This results in an in-plane isotopic concentration gradient during an isotope exchange, which can be followed by performing consecutive measurements at different distances from the open surface, as schematically shown in the lower panel of Fig.~\ref{fig:IERS_schematic}.
The diffusion profiles as a function of time, obtained by interpolation of the transients at each position, can be modelled by Fick's second law to retrieve tracer exchange and in-plane diffusion parameters. 
The time and spatial resolution of this technique enables an unparalleled analysis of tracer exchange and diffusion processes.


\subsubsection*{Theoretical considerations on mass-transport laws}
The isotopic transitions can be analysed within the framework of irreversible thermodynamics to obtain tracer surface oxygen exchange and/or bulk transport parameters, equivalent to the analysis performed in electrical conductivity relaxation \cite{Karthikeyan2008} and time resolved x-ray diffraction measurements \cite{Moreno2013}. 

The oxygen tracer flux through the surface, $J^*$, can be developed as a function of the concentration difference across the interface \cite{Maier2004}:
\begin{equation}\label{eqn:res:surface_crank}
	J^* = \sum_{i=0}^{\infty}k_i^*({^{18}C}_\text{gas}-{^{18}C}_\text{surface}(t))^i,
\end{equation}
with the tracer concentration in the atmosphere, ${^{18}C}_\text{gas}$, the outermost surface layer of the solid, ${^{18}C}_\text{surface}(t)$ and the tracer surface exchange coefficient, $k^*_i$. 
For simple homogenous kinetics and small deviations from equilibrium, its expansion is commonly cut after the linear term \cite{footnote2}.

Using Eq.~\ref{eqn:res:surface_crank} in its linear form and a plane sheet geometry the solution to Fick's second diffusion law in the case of a single rate determining step (RDS) within the surface is given by: 
\begin{equation}\label{eqn:res:Fick}
	\begin{aligned}
		{C'}(t) & = \frac{{^{18}C}(t)-{^{18}C}_\text{ini}}{{^{18}C}_\text{gas}-{^{18}C}_\text{ini}} \propto \frac{\omega(t)-\omega_\text{ini}}{\omega_{\infty}-\omega_\text{ini}} \\
		& = 
		\begin{cases}
			1-\exp^{-t/\tau}. \\
			1-\sin^2{a}\exp^{-t/\tau_\alpha}-\cos^2{a}\exp^{-t/\tau_\beta}
		\end{cases}
	\end{aligned}
\end{equation}
with the normalised isotopic fraction, ${C'}$, the $^{18}$O abundance within the sample at $t=0$ and the gas phase, ${^{18}C}_\text{ini}$ and ${^{18}C}_\text{gas}$, respectively. The characteristic time constant, $\tau$, is linked to the rate-limiting step via $\tau=d_\text{f}/k^*$, with the film thickness, $d_\text{f}$.
However, it is often found that a single exponential decay does not describe sufficiently well the experimental data, which is compensated by introducing a parallel contribution using two time constants, $\tau_\alpha$ and $\tau_\beta$, and a weighting factor, $a$.
This is commonly explained by differences in exposed surface areas, such as grain and grain boundary or different terminations \cite{Chen2003,Kim2006,Yan2019,Stangl2021d}. 
Additionally, we consider the case of a non-linear reaction rate by expanding Eq.~\ref{eqn:res:surface_crank} to the second order. For a surface limited reaction the solution of Fick's law becomes:
\begin{equation}\label{eqn:res:k_expansion_quadr}
	{C'}(t) = 1-\frac{\tau_2}{\left(\tau_1+\tau_2\right)\exp[t/\tau_1]+\tau_1 },
\end{equation}
with $k_i^*=d_\text{f}/\tau_i$.

\begin{figure*}[t]
      	\centering
      	
      	\begin{overpic}[width=175mm, unit=1mm]
      		{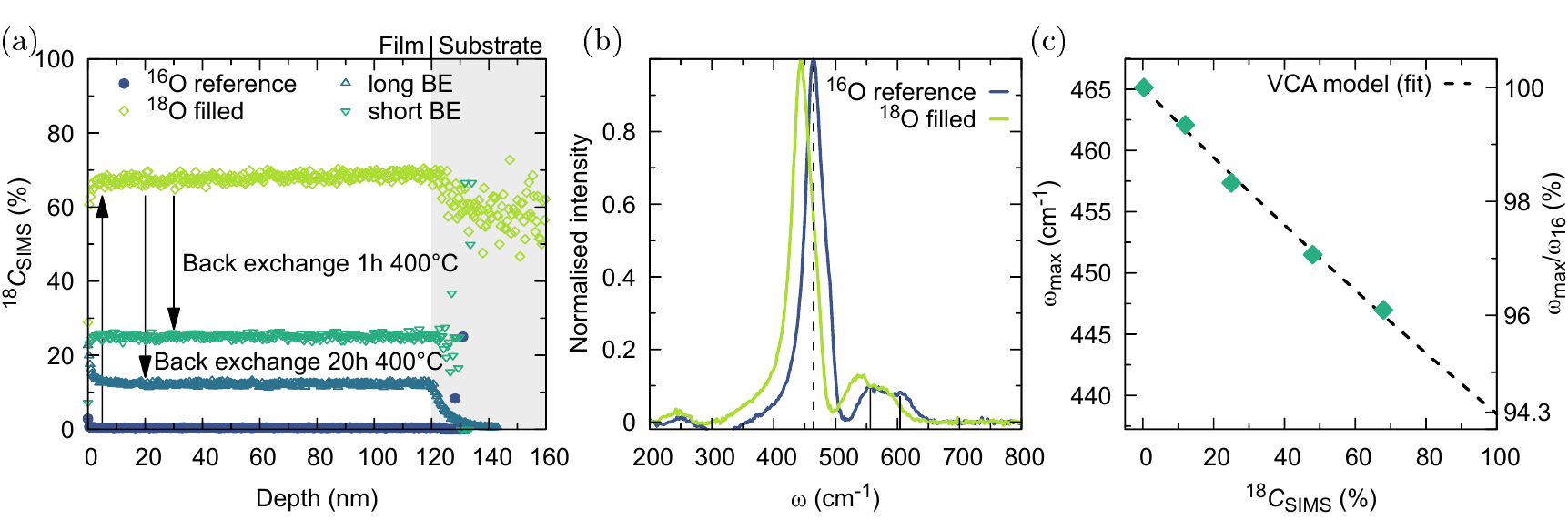}
      	\end{overpic}
      	\caption[]{\textbf{Isotopic Raman shift:} (a) ToF-SIMS $^{18}$O depth profiles for $^{16}$O annealed reference sample, $^{18}$O exchanged film and two films back-exchanged at 400\,°C for 1\, and 20\,h, respectively. (b) $^{18}$O induced shift as observed by Raman spectroscopy. (c) Correlation of Raman shift and the isotopic concentration. The dashed line is the fit of the data to the virtual crystal approximation (VCA).}
      	\label{fig:Analysis_iso_shift}
\end{figure*}

\subsubsection*{Discussion of novel IERS methodology}
Before moving on to the results section, we would like to review the outlined IERS methodology and address several potential issues, such as thermally induced stress, time resolution and laser light related effects. 
Firstly, rapid heating ramps are required to minimize isotopic exchange during the heating and to obtain the best time resolution, which could lead to crack formation and thus damage the oxide surface. This could alter the observed isotope exchange rates, \textit{e.g.} by increasing the exposed surface area, or the exposure of different surface terminations. For accurate measurements the material under investigation must be able to withstand high thermally induced stress. 

Secondly, the spectrum acquisition time must be small compared to the time constant of the exchange process in order to provide sufficient time resolution while ensuring an adequate signal-to-noise ratio and avoiding an artificial line broadening. This requirement defines a material specific upper temperature limit for the applicability of IERS.

Thirdly, laser induced heating and/or photoenhanced activity can potentially affect the thermally activated intrinsic exchange activity of the material.
Photoenhanced activation of oxygen exchange is long known \cite{Formenti1972,Pichat2007} and is understood via the formation of $e'$-\ch[kroeger-vink]{$h$^{.}} pairs across the band gap (for $h\nu>E_\text{gap}$), which increases the electron concentration in the conduction band \cite{Merkle2001,Merkle2002}. Provided that electrons are the rate determining species of oxygen incorporation, irradiation may enhance the surface exchange rates.
Additionally, it was shown recently that ionic conductivity of CGO grain boundaries can be significantly improved upon illumination with above bandgap light \cite{Defferriere2022}.
Material dependent effects of laser light irradiation either due to direct laser heating or photoenhanced activity have to be excluded for accurate assessment of transport kinetics.

\section*{Results and discussion}
\subsection*{Structural and chemical characterisation}

\begin{figure*}[t]
	\centering
	
	\begin{overpic}[width=150mm, unit=1mm]
		{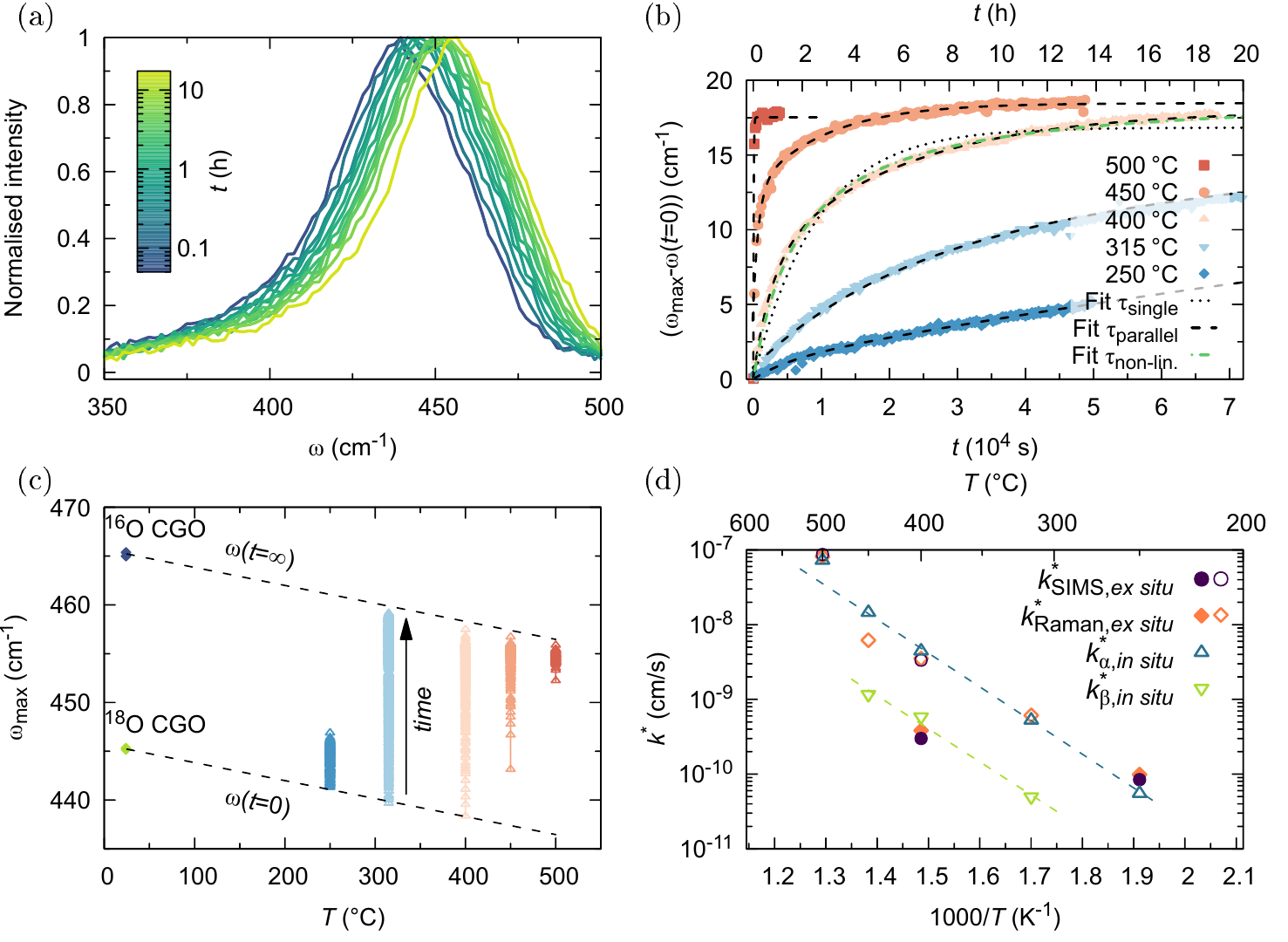}
	\end{overpic}
	\caption[]{\textbf{\textit{In situ} isotopic exchange Raman spectroscopy:} (a) Raman spectrum of CGO as a function of time at 400\,°C. (b) \textit{In situ} back-exchange transients at different temperatures in flowing dry air ($p$O$_2=0.21$\,atm). Dotted, dashed and dashed-dotted lines through the 400\,°C data set correspond to single and double exponential fittings ($\tau_\text{single}$ and $\tau_\text{parallel}$) and the non-linear reaction rate model ($\tau_\text{non-lin.}$). (c) Shift from initial ($\omega(t=0)$) $^{18}$O to (expected) final $^{16}$O state in a ($\omega_\text{max}$-$T$) diagram. (d) Arrhenius-type plot of the surface exchange coefficient, $k^*$, obtained by \textit{in situ} and \textit{ex situ} techniques. Open/closed circles and diamonds correspond to short/long annealings, see main text for details.}
	\label{fig:insitu_Raman}
\end{figure*}
\begin{figure*}[!hb]
	\centering
	\begin{overpic}[width=170mm, unit=1mm]
		{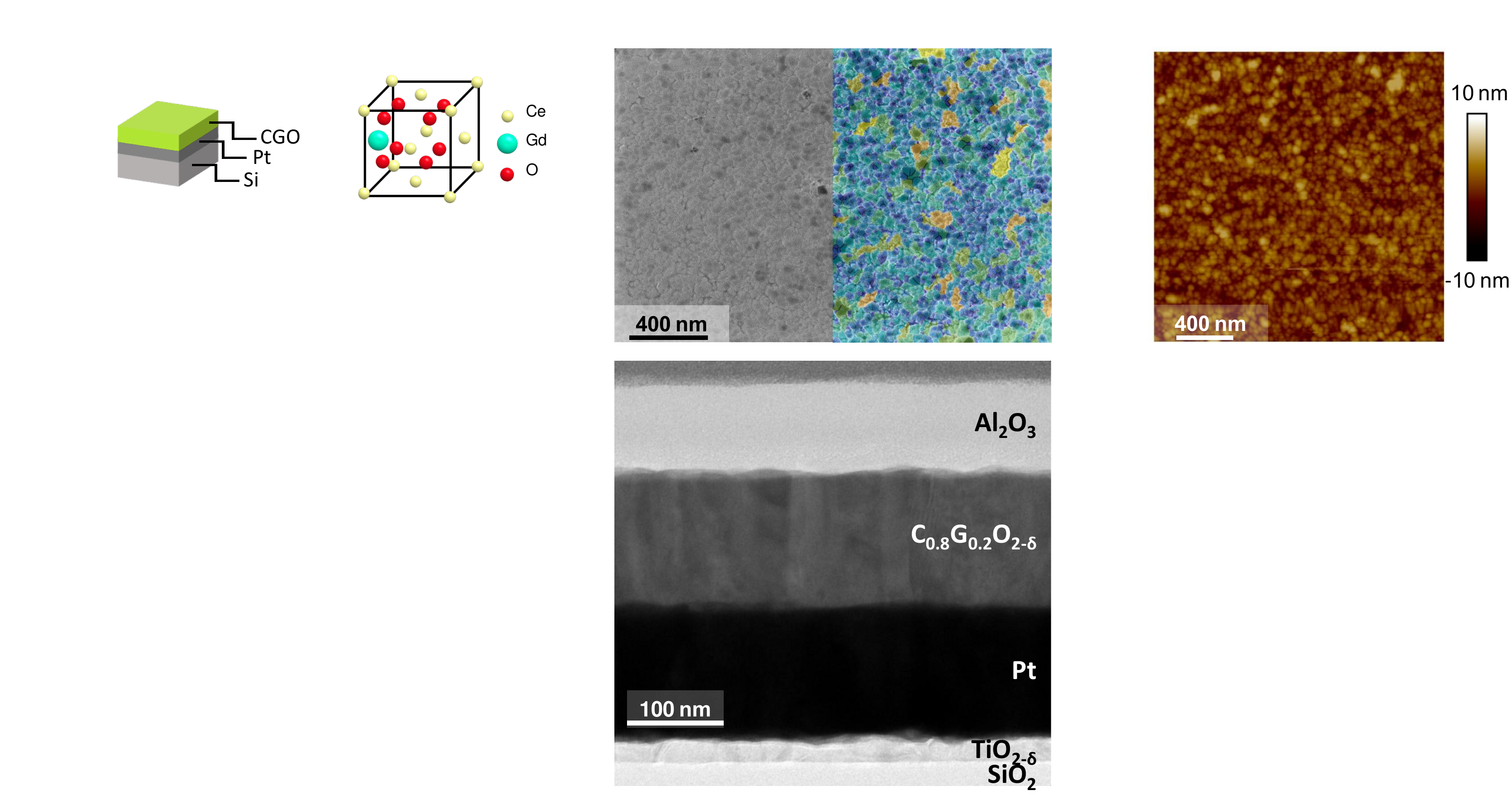}
		\put(0,0){\includegraphics[width=170mm, unit=1mm]{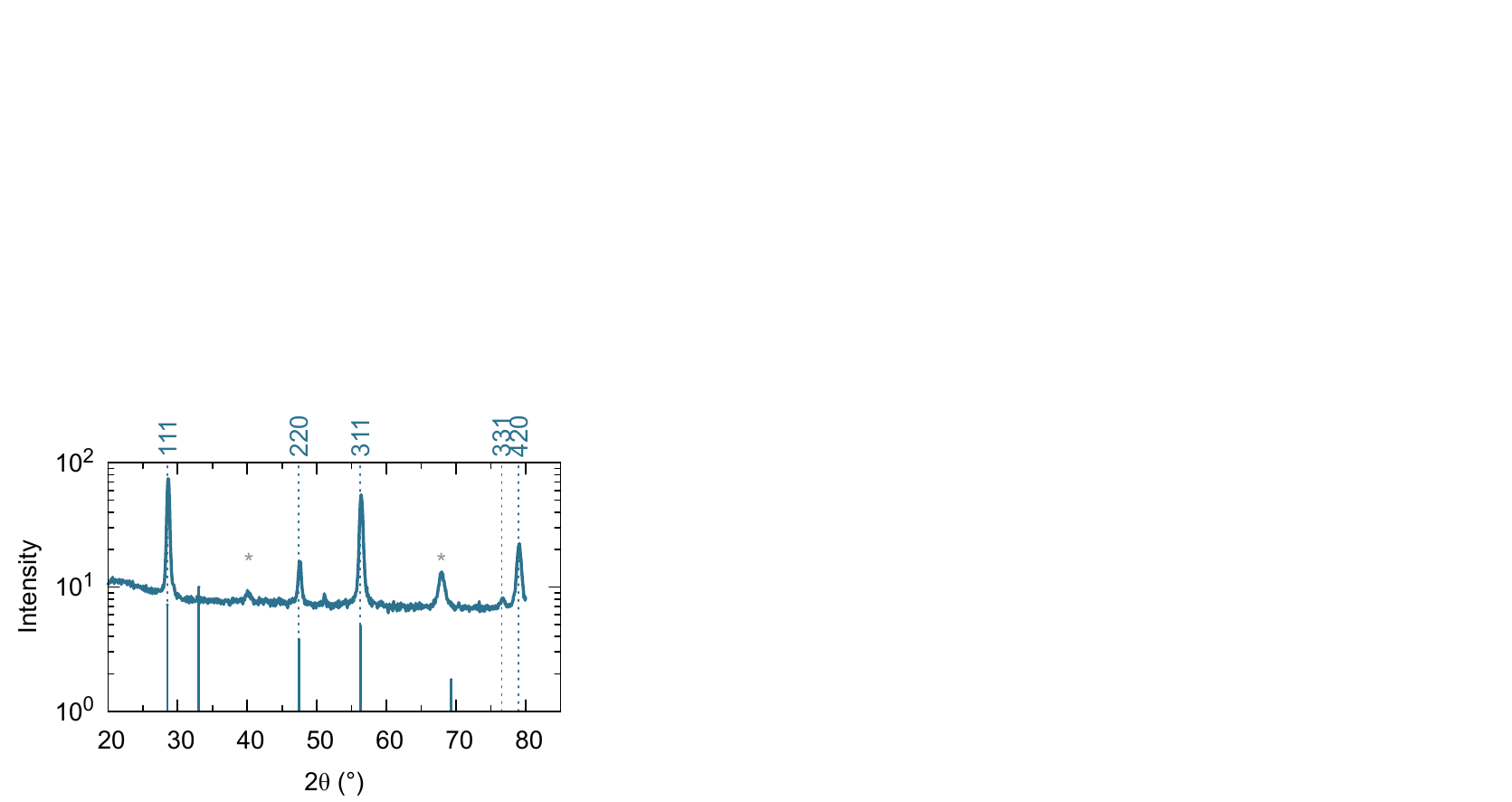}}
		\put(0,0){\includegraphics[width=170mm, unit=1mm]{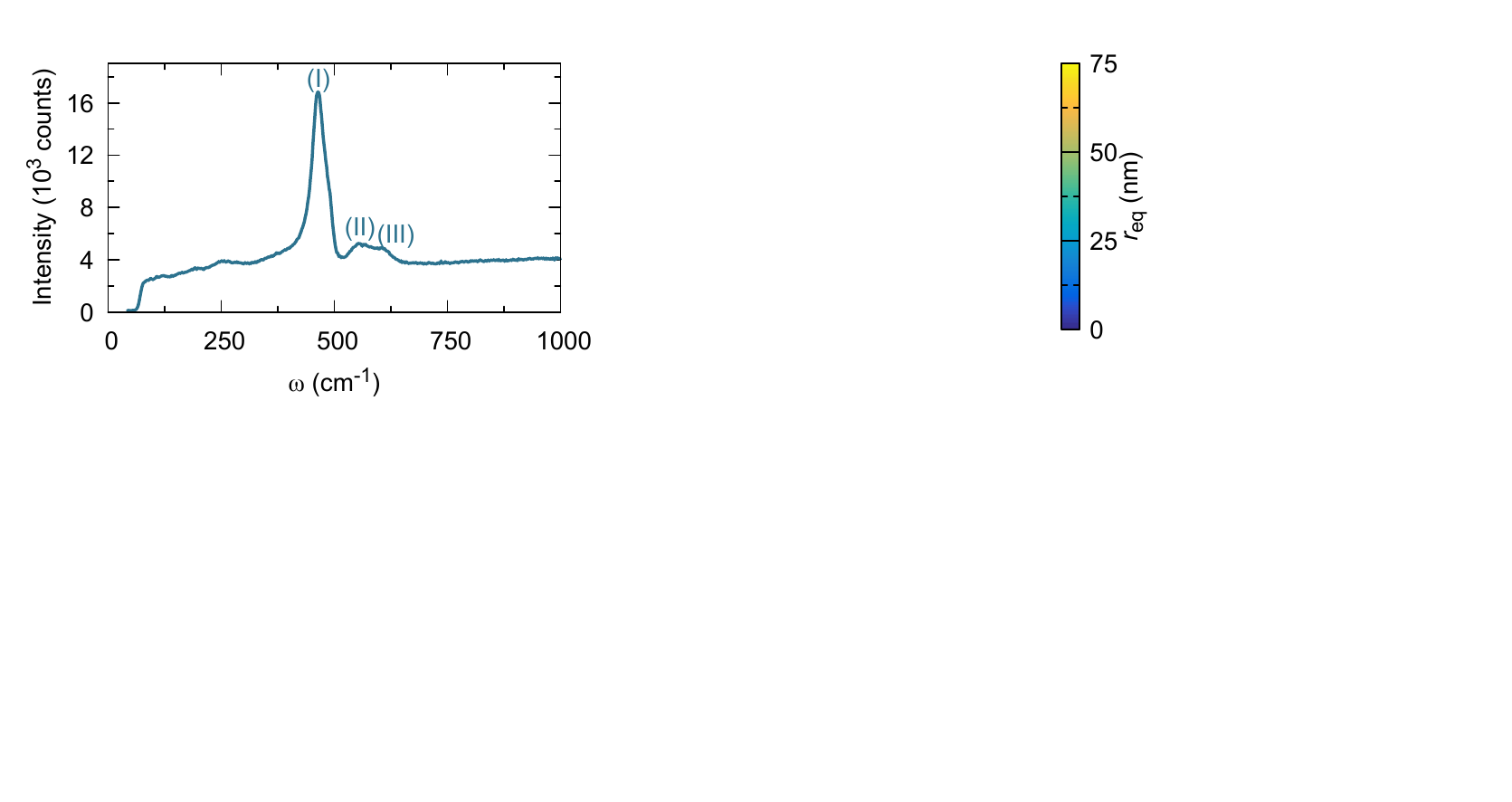}}
		\put(0,0){\includegraphics[width=170mm, unit=1mm]{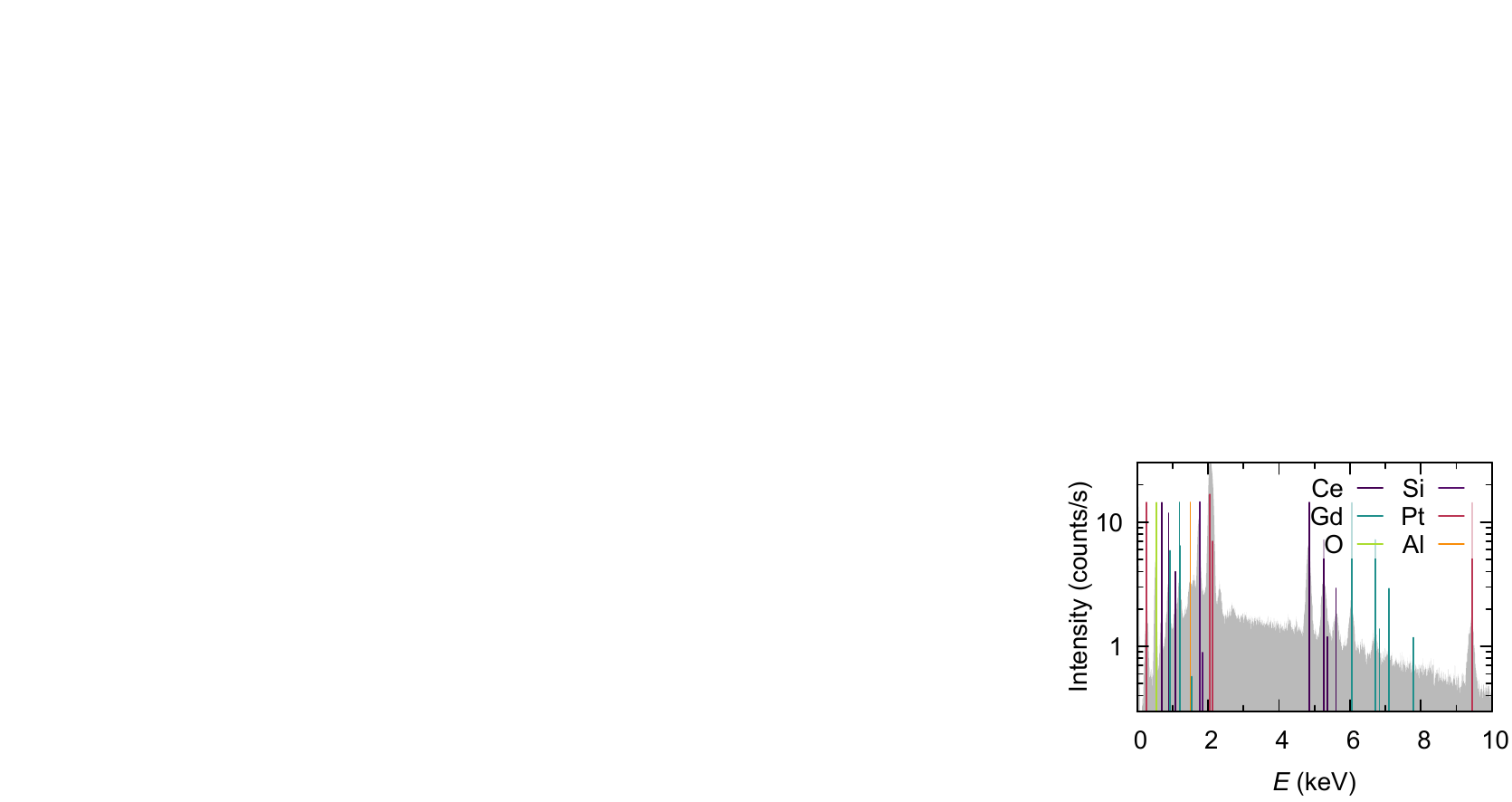}}
		\put(2,86){(a)}
		\put(2,44){(b)}
		\put(63,86){(c)}
		\put(63,44){(d)}
		\put(129,86){(e)}
		\put(119,44){(f)}
	\end{overpic}
	\caption[]{\textbf{Material characterisation of the CGO/Pt/Si thin films:} (a) Raman spectrum of CGO composed of the oxygen breathing mode (I) and defect modes (II \& III) induced via Gd substitution. Inset shows schematic of sample architecture and CGO crystal structure. (b) Grazing incidence X-ray diffraction pattern showing textured growth. Films are free of any secondary phases. Substrate peaks at 40.2 and 67.9\,° are marked with asterisks. (c) Secondary electron SEM top view image reveals a homogenous, dense structure, with nanometric grain sizes of $48\pm23$\,nm. False colour plot shows equivalent grain radius, $r_\text{eq}$. The corresponding distribution of the grain sizes is shown in Fig.~\ref{fig:SI_fitting}(b), along with its area distribution. (d) Cross-section transmission electron microscopy (TEM) image of 120\,nm CGO film on top of Pt with 90\,nm alumina coating. (e) Topographical atomic force microscopy (AFM) surface map. (f) Energy dispersive X-ray spectrum of a CGO sample.}
	\label{fig:SI_Material_analysis}
\end{figure*}

Polycrystalline Ce$_{0.8}$Gd$_{0.2}$O$_{2-\delta}$ thin films were deposited by PLD on Pt/Si substrates (see experimental for more details). The film thicknesses of the studied samples were about 120\,nm with a dense structure, nanometric grains ($48\pm23$\,nm) and low roughness ($R_\text{q} = 1.6$\,nm). A summary of their main characteristics is shown in Fig.~\ref{fig:SI_Material_analysis}.

Cerium oxide has the cubic fluorite structure (O$_\text{h}^5$ (Fm-3m) space group). This structure has six optical-phonon branches, which yield one infrared active phonon of T$_\text{1u}$ symmetry and one Raman active phonon of T$_\text{2g}$ symmetry at $k=O$. The infrared and Raman active modes involve the anti-symmetric and symmetric stretching vibrations of the cation-oxygen O-Ce-O units, respectively. The Raman-active mode therefore only involves the motion of oxygen atoms. For undoped cerium oxide, this mode is observed at approximately 465\,cm$^{-1}$. 

The Raman spectrum of an as deposited CGO thin film with natural $^{18}$O isotopic abundance (0.2\,\%) is shown in Fig.~\ref{fig:SI_Material_analysis}(a) (see also Fig.~\ref{fig:Analysis_iso_shift}(b)).
While its main band still arises around 465\,cm$^{-1}$, doping with trivalent rare earth cations, like Gd$^{3+}$ creates a symmetry breaking in the lattice, resulting in a long tail evolving on the left shoulder of the band \cite{McBride1994} and two additional features between 550 and 600\,cm$^{-1}$ in the Raman spectrum of CGO. 
These additional bands are disorder-induced modes related to oxygen vacancies created to maintain charge neutrality when Gd$^{3+}$ ions are incorporated in Ce$^{4+}$ sites, according to $2$\ch[kroeger-vink]{[Gd_{Ce}^{'}]}=\ch[kroeger-vink]{[V_{O}^{..}]} \cite{Yang2016}.
The contribution of the Si substrate is effectively blocked by the metallic Pt buffer layer. The T$_\text{2g}$ mode position is obtained by line fitting, as exemplarily shown in Fig.~\ref{fig:SI_fitting}(a).

\begin{figure*}[]
	\centering
	\begin{minipage}[b]{0.49\textwidth}
		\begin{overpic}[width=\textwidth, unit=1mm]
			{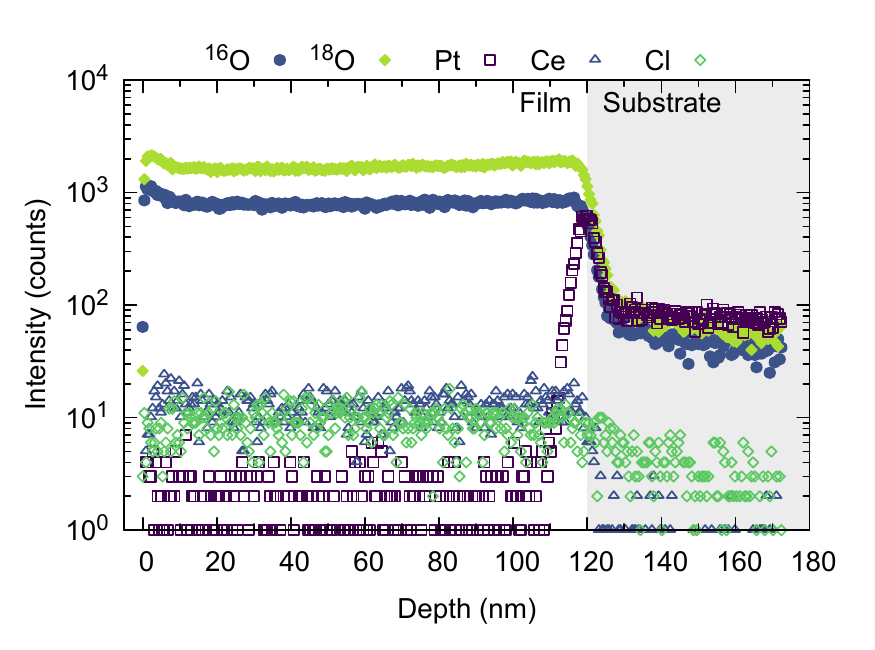}
			\put(-2.5,51){(a)}
		\end{overpic}
	\end{minipage}
	\begin{minipage}[b]{0.49\textwidth}
		\begin{overpic}[width=\textwidth, unit=1mm]
			{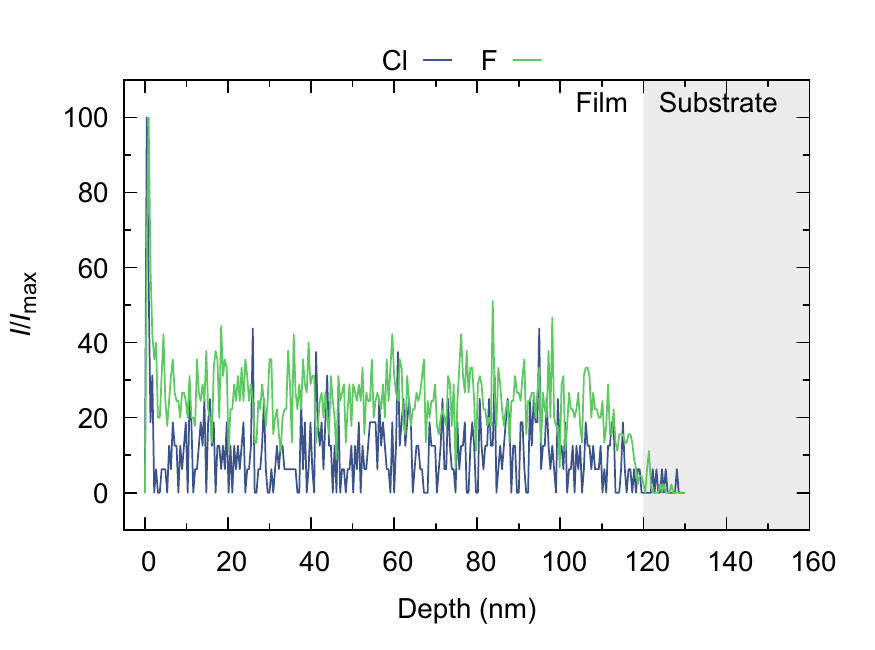}
			\put(-2.5,51){(b)}
		\end{overpic}
	\end{minipage}
	\caption[]{SIMS ion depth profiles through a 120\,nm thick CGO film on Pt substrate exchanged at 640\,°C for 0.5\,h: (a) Selected film and substrate secondary ions and (b) Cl$^-$ and F$^-$ surface contaminants (normalised to maximum value). The steep drop of surface contaminant species within the first few nm confirms a good depth resolution without ion-beam mixing.}
	\label{fig:SI:SIMS_CGO_Pt}
\end{figure*}

Subsequently, some of the CGO films were enriched in $^{18}$O.
The elemental and isotopic concentration as a function of film thickness was analysed by ToF-SIMS for several samples at different stages of $^{18}$O enrichment, as shown in Fig.~\ref{fig:Analysis_iso_shift}(a) for a CGO sample exchanged at 640\,°C (0.5\,h) and two samples partially $^{16}$O back-exchanged at 400\,°C (1\,h and 20\,h, respectively). 
For all studied films, the resulting ${^{18}C}$ curves are flat. The absence of a diffusion profile indicates that the oxygen exchange is limited by surface processes while bulk diffusion is comparably fast. 
The depth profiles of selected secondary ion species are shown in Fig.~\ref{fig:SI:SIMS_CGO_Pt} for a CGO sample after $^{18}$O exchange at 640\,°C.

The substitution of heavy oxygen into the crystal lattice effectively leads to a modification of the vibrational spectra and results in a red shift of the T$_\text{2g}$ mode of about 20\,cm$^{-1}$, \textit{i.e.} 4.3\,\%, between the $^{16}$O reference and $^{18}$O exchanged CGO sample, as shown in Fig.~\ref{fig:Analysis_iso_shift}(b). The strong downshift of the defect-induced lines upon $^{18}$O substitution confirms that these modes involve oxygen motion as well.

\begin{figure}[t]
	\centering
	\begin{overpic}
		{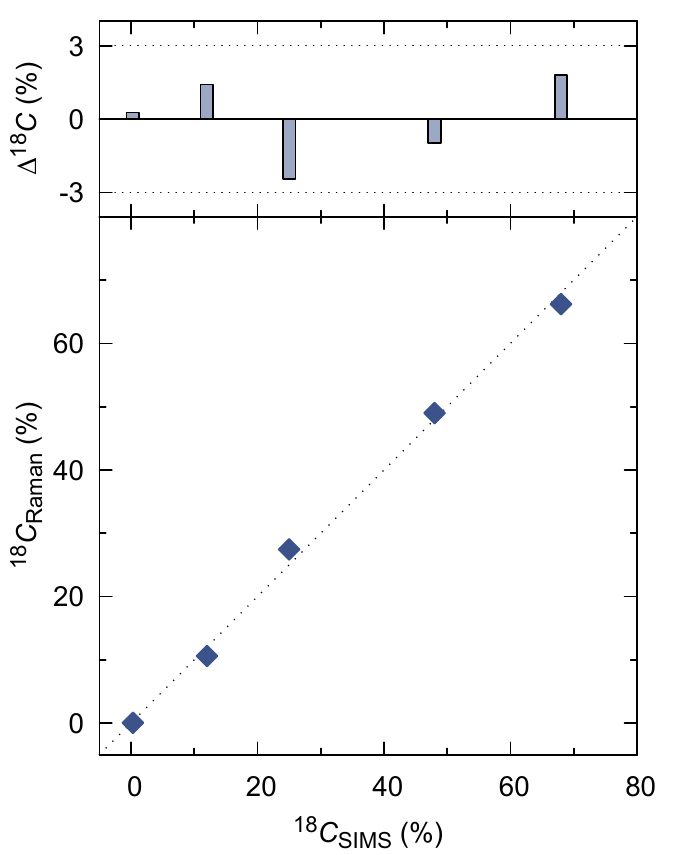}
	\end{overpic}
	\caption[]{Comparison of $^{18}$O isotopic fraction, ${^{18}C}$, obtained from Raman measurements via VCA model \textit{vs.} SIMS values and the resulting mismatch (upper panel).}
	\label{fig:SI_Raman_sims_compare}
\end{figure}

The dependence of the maximum position of the T$_\text{2g}$ mode, $\omega_\text{max}$, on the isotopic concentration is shown in Fig.~\ref{fig:Analysis_iso_shift}(c), with the $^{18}$O filling level measured by ToF-SIMS.
The dashed line shows the best fit of experimental data using the VCA model, as given by Eq.~\ref{eqn:Intro:VCA_fit}. 
The observed line shifts correspond to those expected, within the experimental accuracy. Thus, ${^{18}C}$ can be precisely assessed via Raman spectroscopy, see also Fig.~\ref{fig:SI_Raman_sims_compare}.

It is worthwhile to note that in off-resonant conditions (as is the case here) the penetration depth of the laser exceeds the film thickness. The isotopic concentration, obtained via the measured Raman shift, therefore corresponds to an average concentration across the film thickness. Any inhomogeneous $^{18}$O concentration along the out of plane direction is expected to lead to a modification of the line shape of the T$_\text{2g}$ mode, \textit{i.e.} the peak width and profile. This is not observed in the present films and a homogenous oxygen concentration can be assumed, in agreement with flat SIMS depth profiles. Thus, the T$_\text{2g}$ mode frequency is the only information needed to determine the $^{18}$O content in the thin films. 
In addition, the grain size is much smaller than the size of the probing laser spot (1\,µm$^2$), which means that different diffusion paths, \textit{e.g.} within the grains and along the grain boundaries, will not be distinguished.

\begin{figure*}[t]
	\centering
	
	\begin{overpic}[width=150mm, unit=1mm]
		{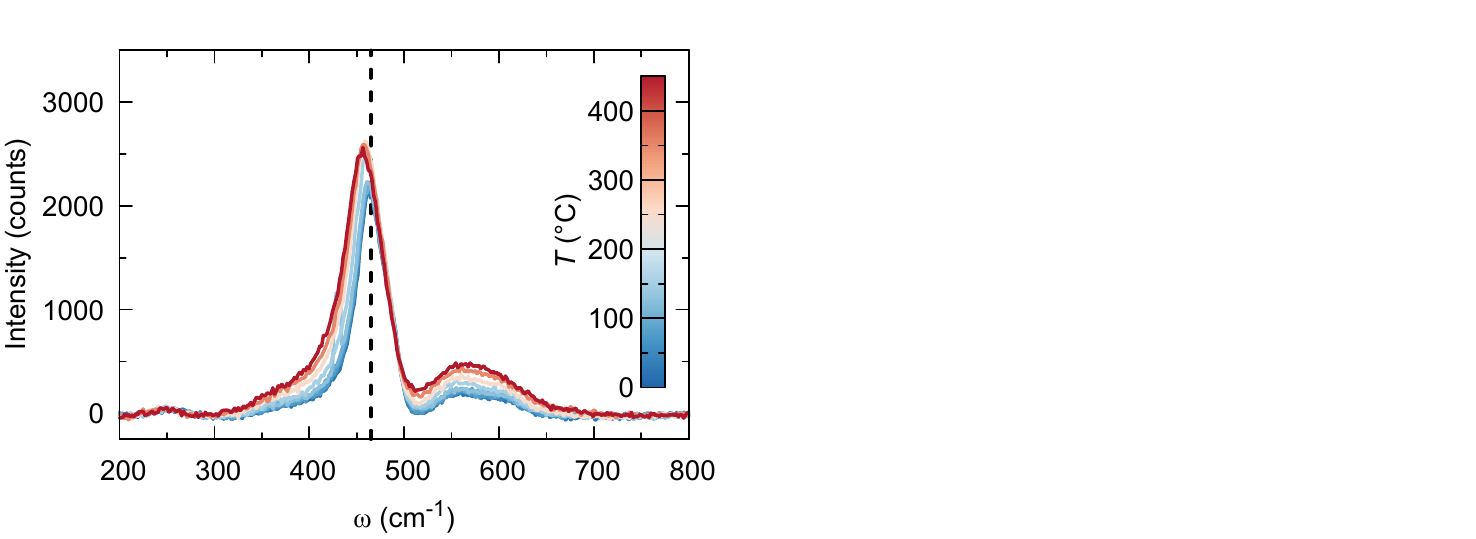}
		\put(0,0){\includegraphics[width=150mm, unit=1mm]{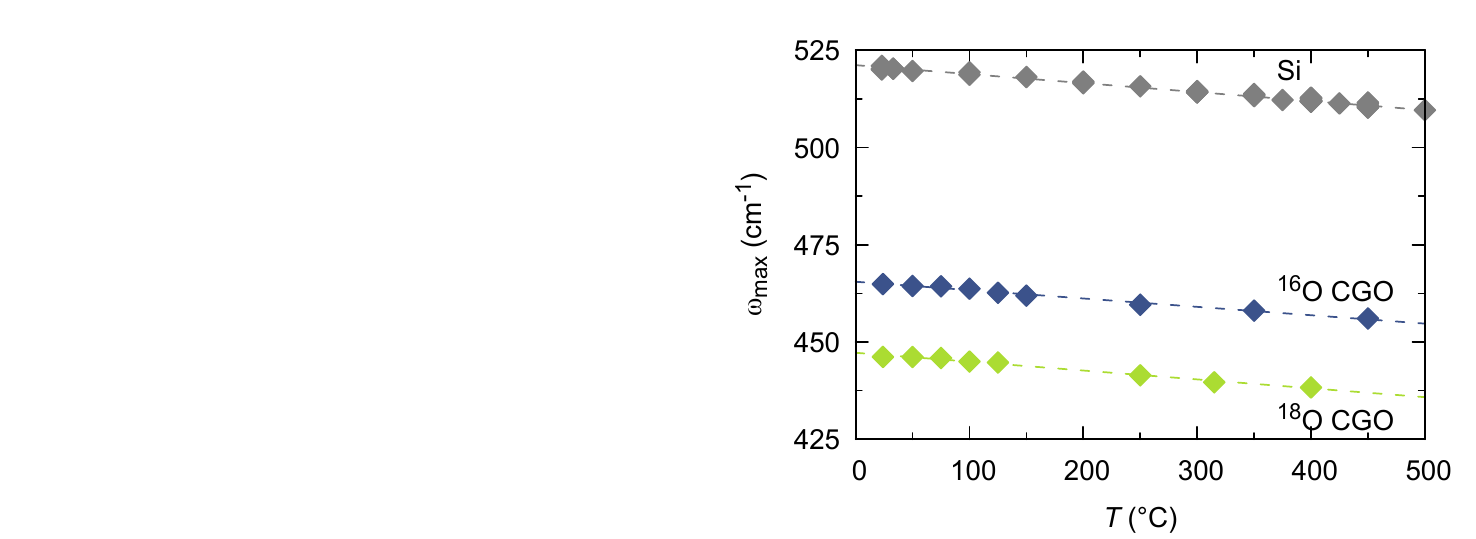}}
		\put(2,52){(a)}
		\put(75,52){(b)}
	\end{overpic}
	\caption[]{Temperature evolution of CGO Raman spectra: (a) Baseline corrected Raman spectra of CGO with natural $^{18}$O composition at temperatures between 22\,°C to 450\,°C. The room temperature position is indicated by the vertical dashed line. (b) Shift of $\omega_\text{max}$ with temperature for $^{16}$O reference and $^{18}$O enriched CGO films, and for a Si reference sample. Dashed lines are guide to the eye. Note that all points shown here were obtained in dry air atmosphere with natural $^{16}$O abundance. However, the high temperature points for the $^{18}$O sample are the first points during back-exchange experiments and therefore they are out of equilibrium. For temperatures higher than the ones shown, kinetics are too fast and the first measured points do not correspond to the initial $^{18}$O state at $t=0$, as discussed in the main text.}
	\label{fig:SI:Raman_T_calibration}
\end{figure*}

Finally, the evolution of the Raman spectra with temperature is shown in Fig.~\ref{fig:SI:Raman_T_calibration}(a). The T$_\text{2g}$ mode remains well defined at elevated temperatures, while thermal expansion of the lattice and anharmonic effects both induce a quite linear shift of its maximum position to lower wavenumbers, as depicted in Fig.~\ref{fig:SI:Raman_T_calibration}(b), accompanied by a broadening of the mode. The frequency of the T$_\text{2g}$ mode of both, $^{16}$O reference and $^{18}$O enriched films exhibit a very similar temperature dependence, \textit{i.e.} the isotopic shift, $\Delta\omega$, remains approximately constant with temperature. Combining Eq.~\ref{eqn:Intro:VCA_fit} with the linear temperature shift, hence allows one to obtain \textit{in situ} the ${^{18}C}$ concentration at elevated temperatures based on the measurement of Raman frequencies.

%
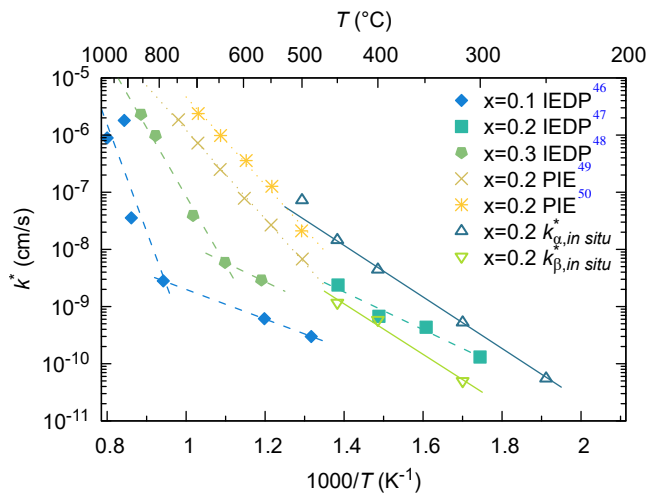
\begin{figure}[t]
	\centering
	\begin{overpic}[width=89mm, unit=1mm]
		{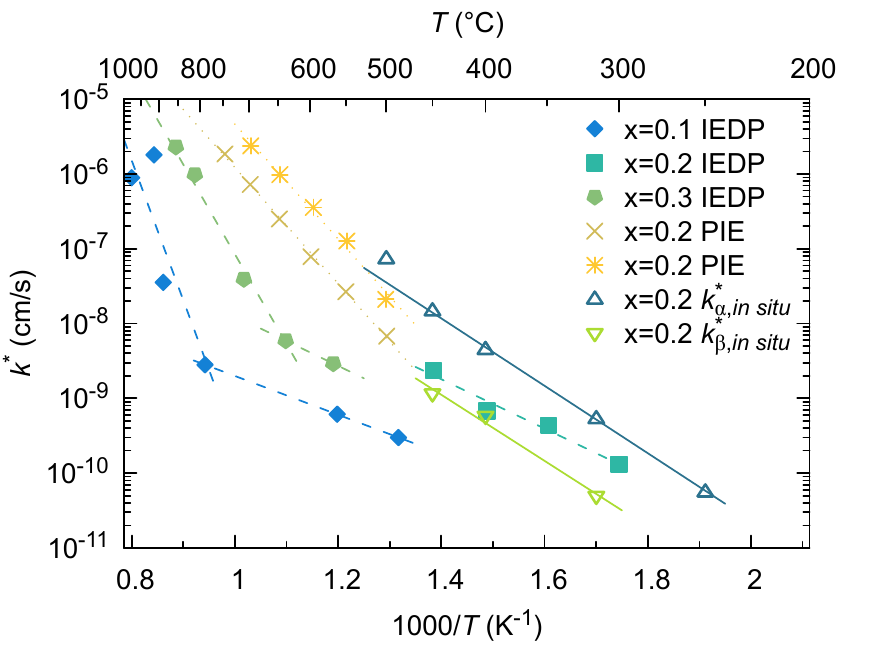}
		\put(78,54){\footnotesize \textsuperscript{\cite{Manning1996}}}
		\put(78,50.5){\footnotesize \textsuperscript{\cite{Kowalski2009}}}
		\put(78,47){\footnotesize \textsuperscript{\cite{Ruiz-Trejo1998}}}
		\put(76,43.5){\footnotesize \textsuperscript{\cite{Schaube2019}}}
		\put(76,40){\footnotesize \textsuperscript{\cite{Yoo2012}}}
	\end{overpic}
	
	\caption[]{\textbf{CGO surface exchange kinetics}: Tracer surface exchange coefficients as reported in literature for different types of Ce$_{1-x}$Gd$_x$O$_2$ samples obtained by IEDP or PIE, compared to \textit{in situ} values found in this work: polycrystalline ceramics ($x=0.1$ \& $x=0.2$ IEDP \cite{Manning1996,Kowalski2009}), single crystal CGO ($x=0.3$ IEDP \cite{Ruiz-Trejo1998}) and CGO particles ($x=0.2$ PIE \cite{Schaube2019,Yoo2012}).}
	\label{fig:CGO_k_literature}
\end{figure}

\subsection*{IERS for surface limited thin film scenario}

\begin{figure*}[]
	\centering
	\begin{overpic}[width=175mm, unit=1mm]
		{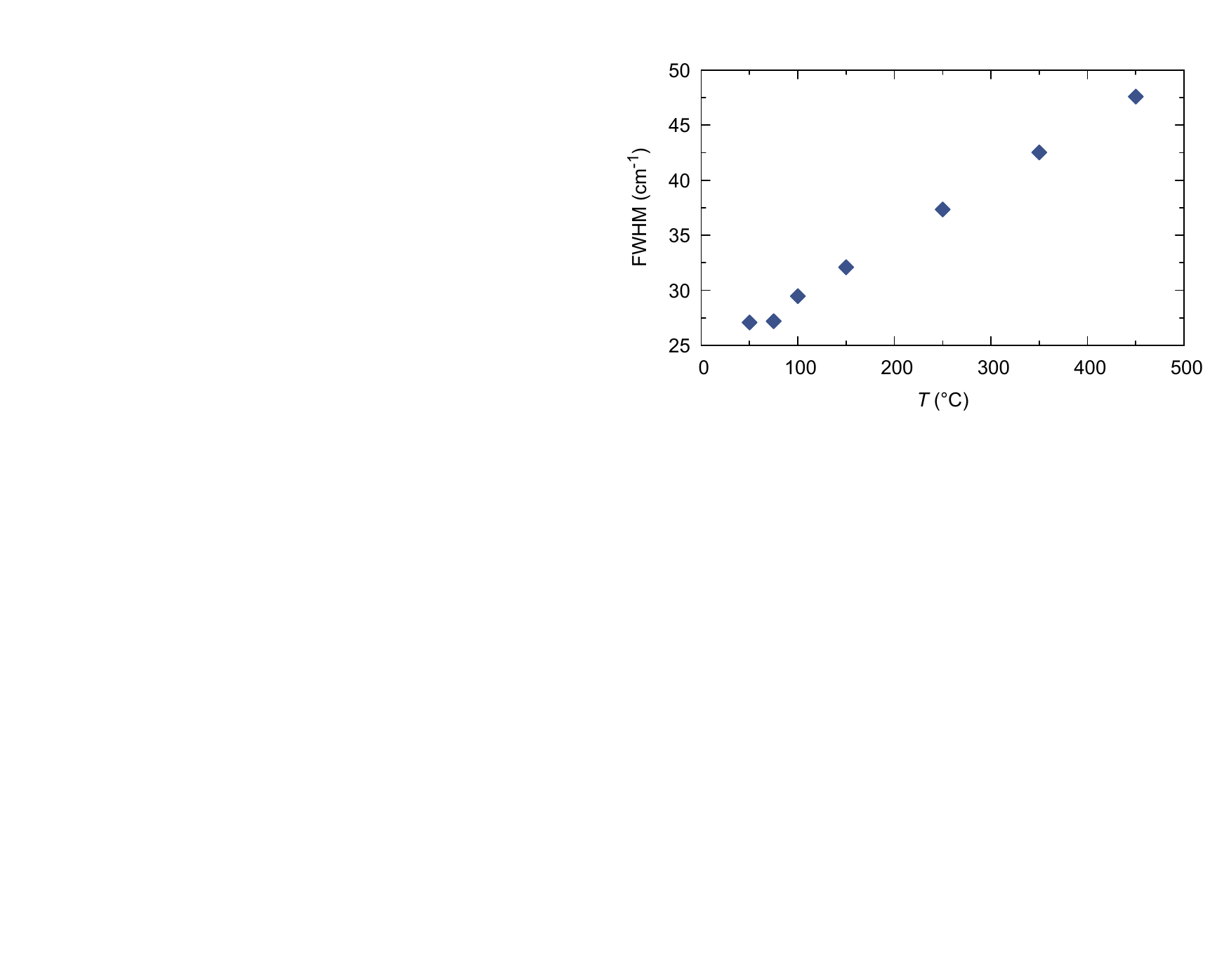}
		\put(0,0){\includegraphics[width=175mm, unit=1mm]{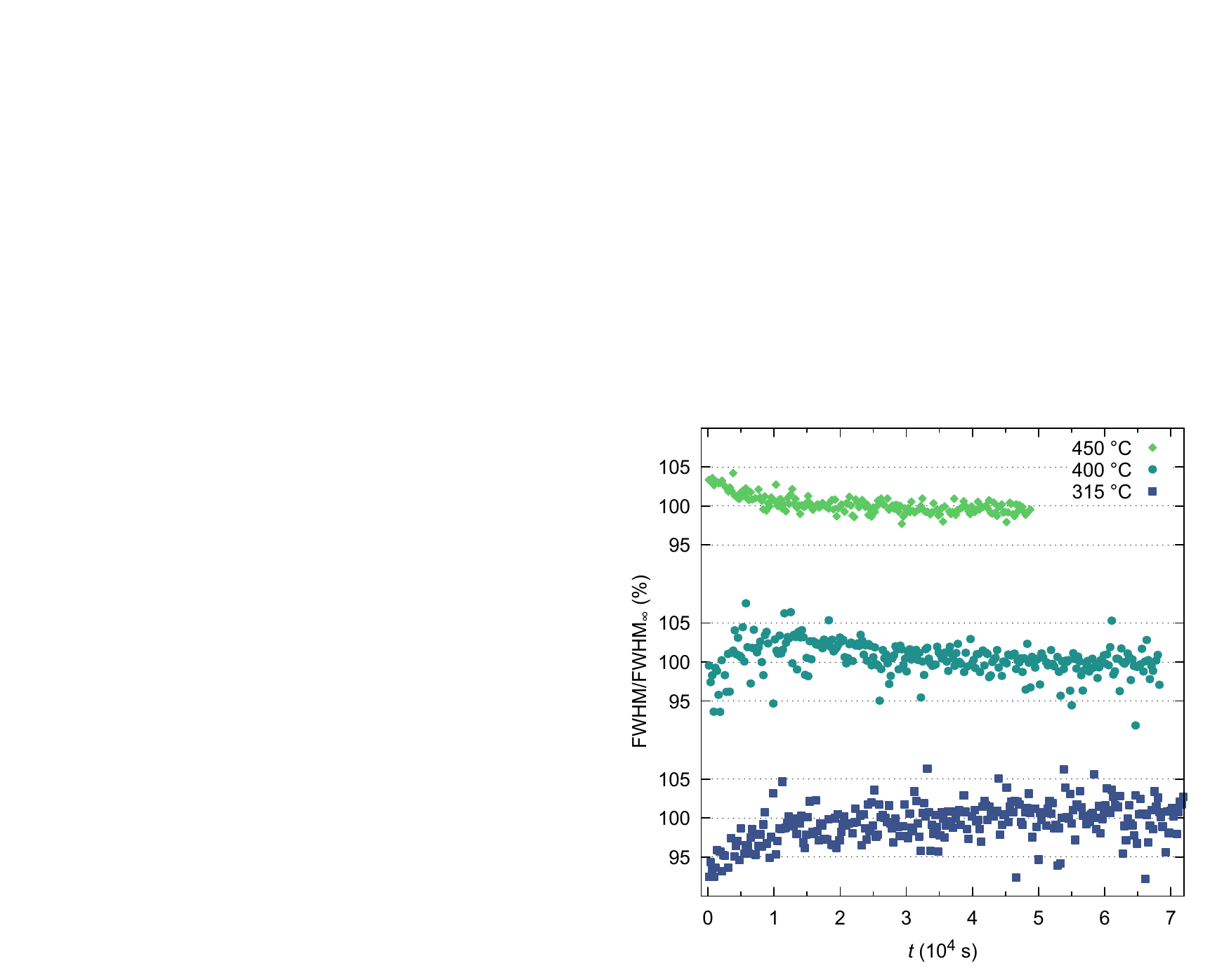}}
		\put(0,0){\includegraphics[width=89mm, unit=1mm]{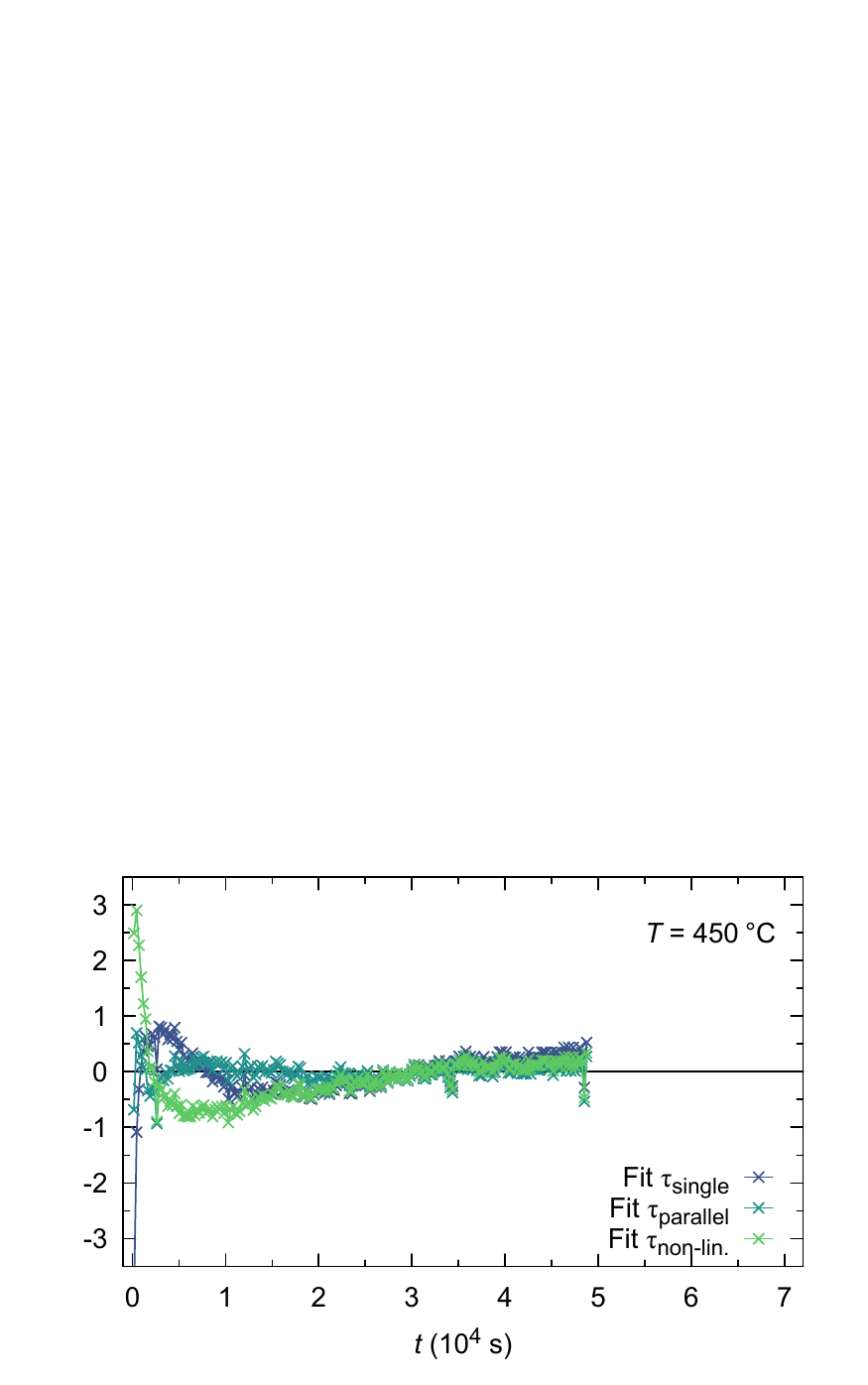}}
		\put(0,0){\includegraphics[width=89mm, unit=1mm]{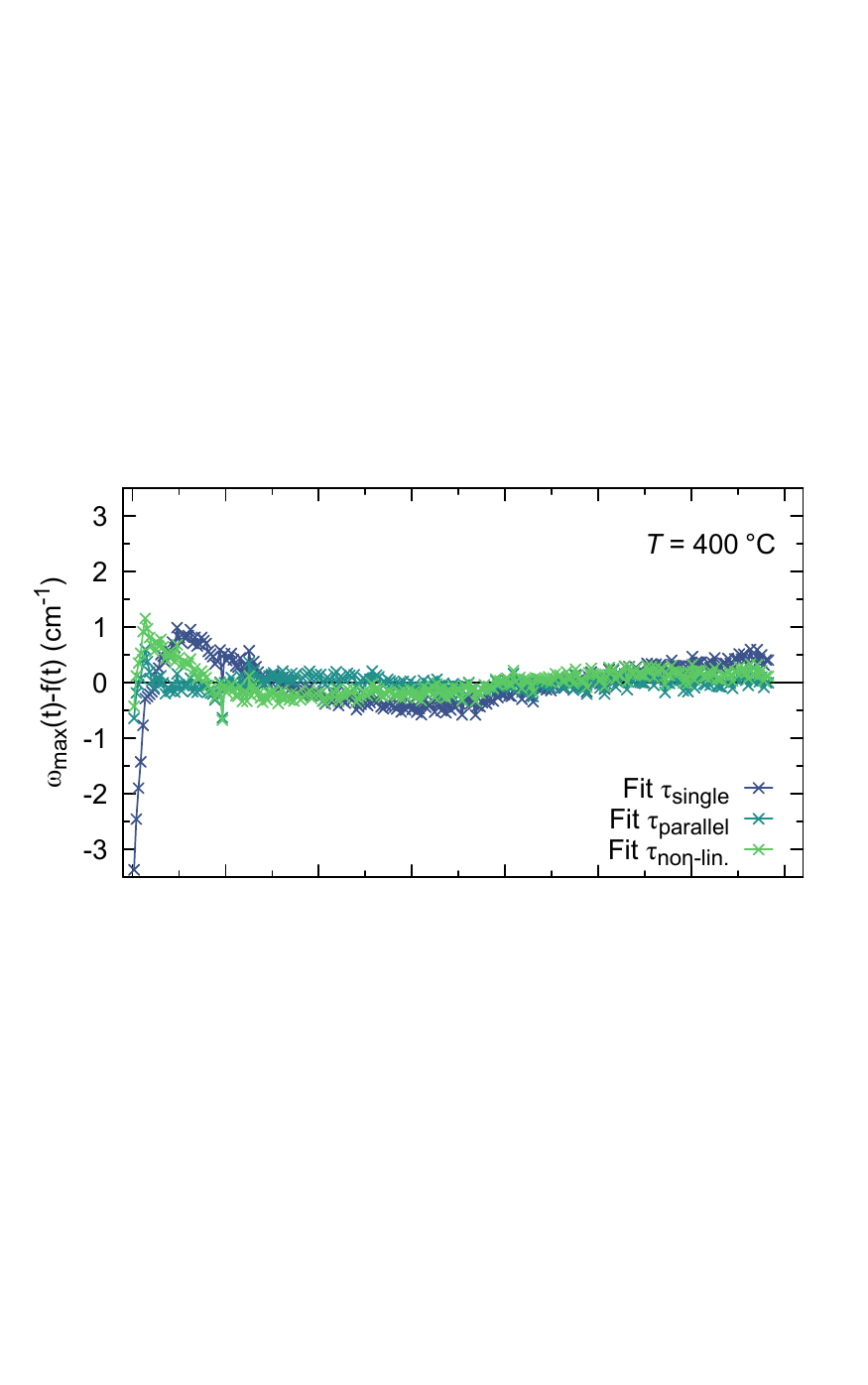}}
		\put(0,0){\includegraphics[width=89mm, unit=1mm]{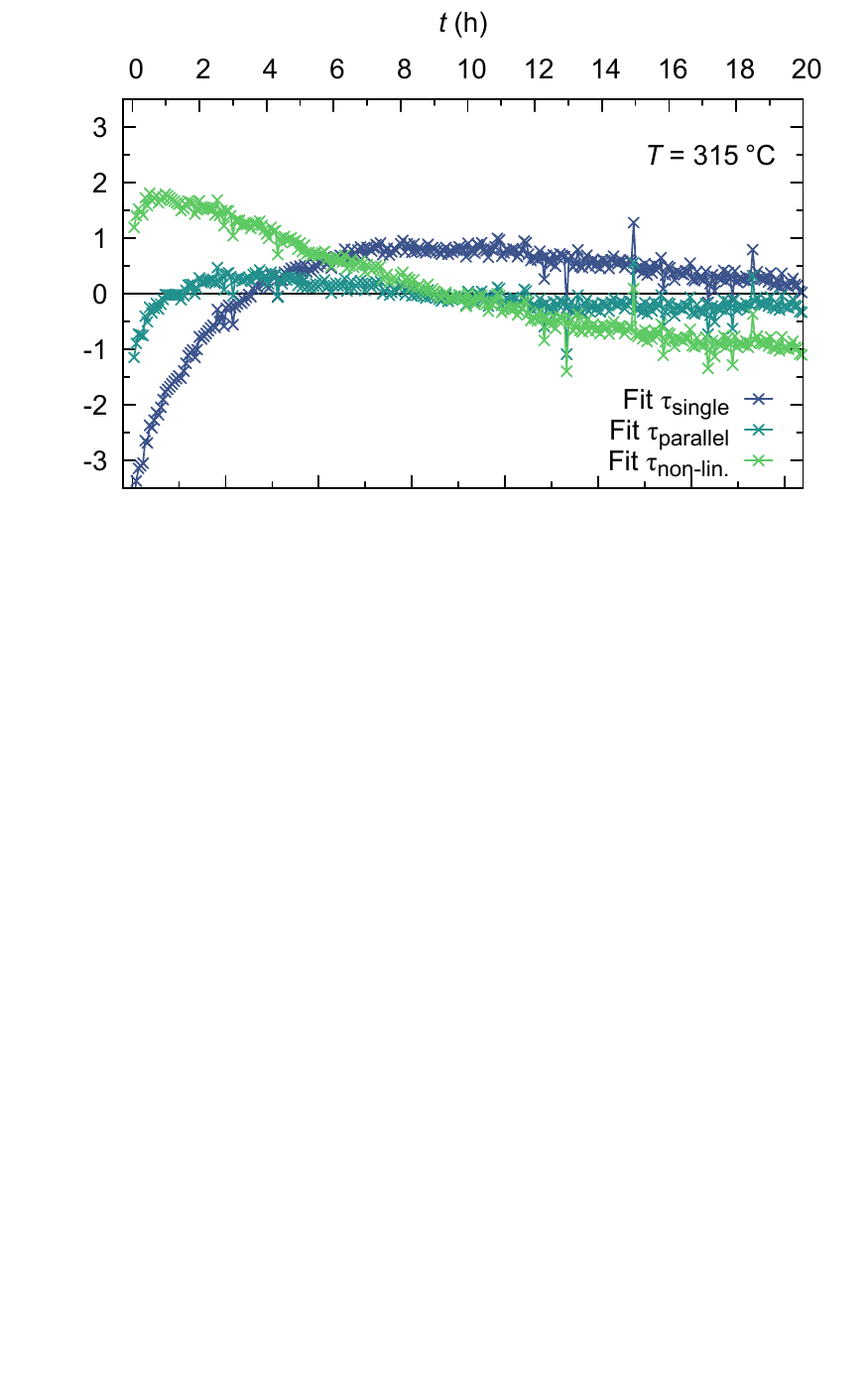}}
		\put(2,132){(a)}
		\put(2,92){(b)}	
		\put(2,52){(c)}
		\put(91,132){(d)}
		\put(91,80){(e)}
	\end{overpic}
	\caption[]{Deviation of \textit{in situ} back-exchange transient from different fitting models at (a) 315\,°C, (b) 400\,°C and (c) 450\,°C: classic linear approach with one ($\tau_\text{single}$) or two parallel exponential decays ($\tau_\text{parallel}$) and model based on non-linear reaction rate ($\tau_\text{non-lin.}$). (d) Evolution of peak width (FWHM) of the T$_\text{2g}$ mode with temperature and (e) time during back-exchange experiments.}
	\label{fig:SI:devation_fit}
\end{figure*}

\begin{figure*}[h]
	\centering
	\includegraphics{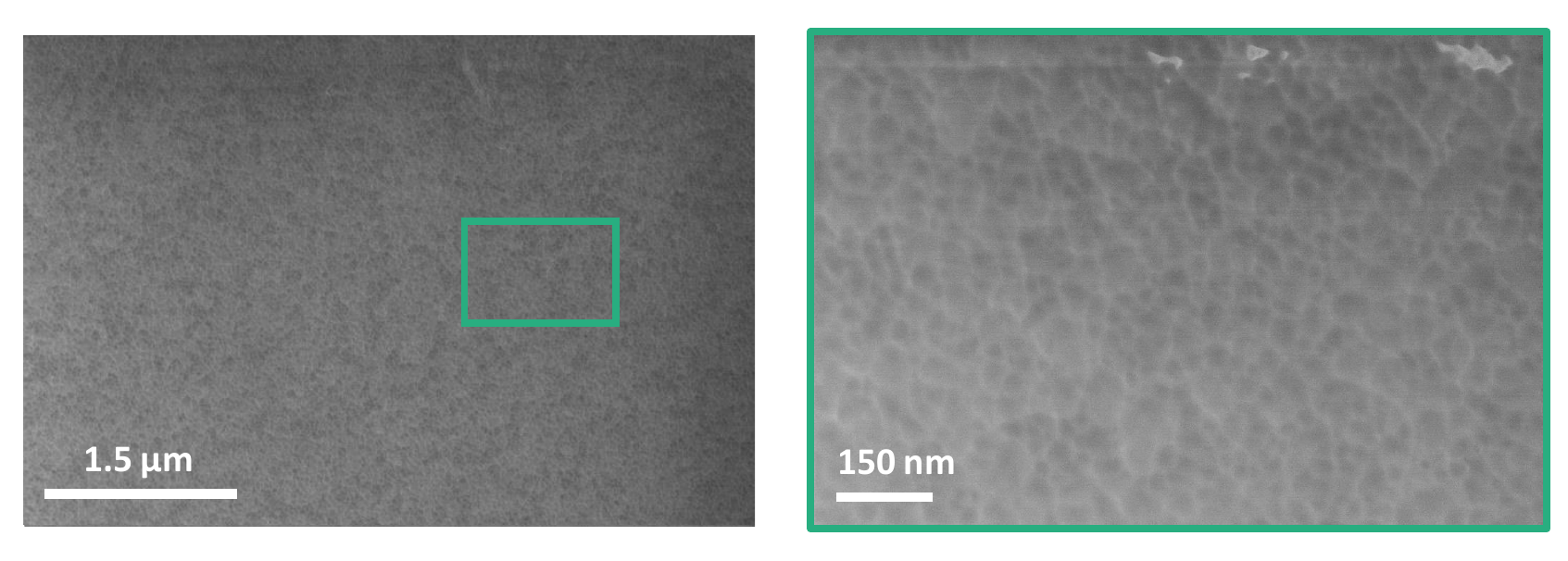}
	\caption[]{SEM images of a CGO thin film after a temperature cycle up to 500\,°C with very fast heating ramps of 100\,°C/min. No formation of cracks is observed, neither on a macroscopic scale, nor on a nanometric scale in the the magnified image on the right.}
	\label{fig:SI:SEM_no_crack}
\end{figure*}

\begin{figure*}
	\centering
	\begin{minipage}[b]{0.49\textwidth}
		\begin{overpic}[width=\textwidth, unit=1mm]
			{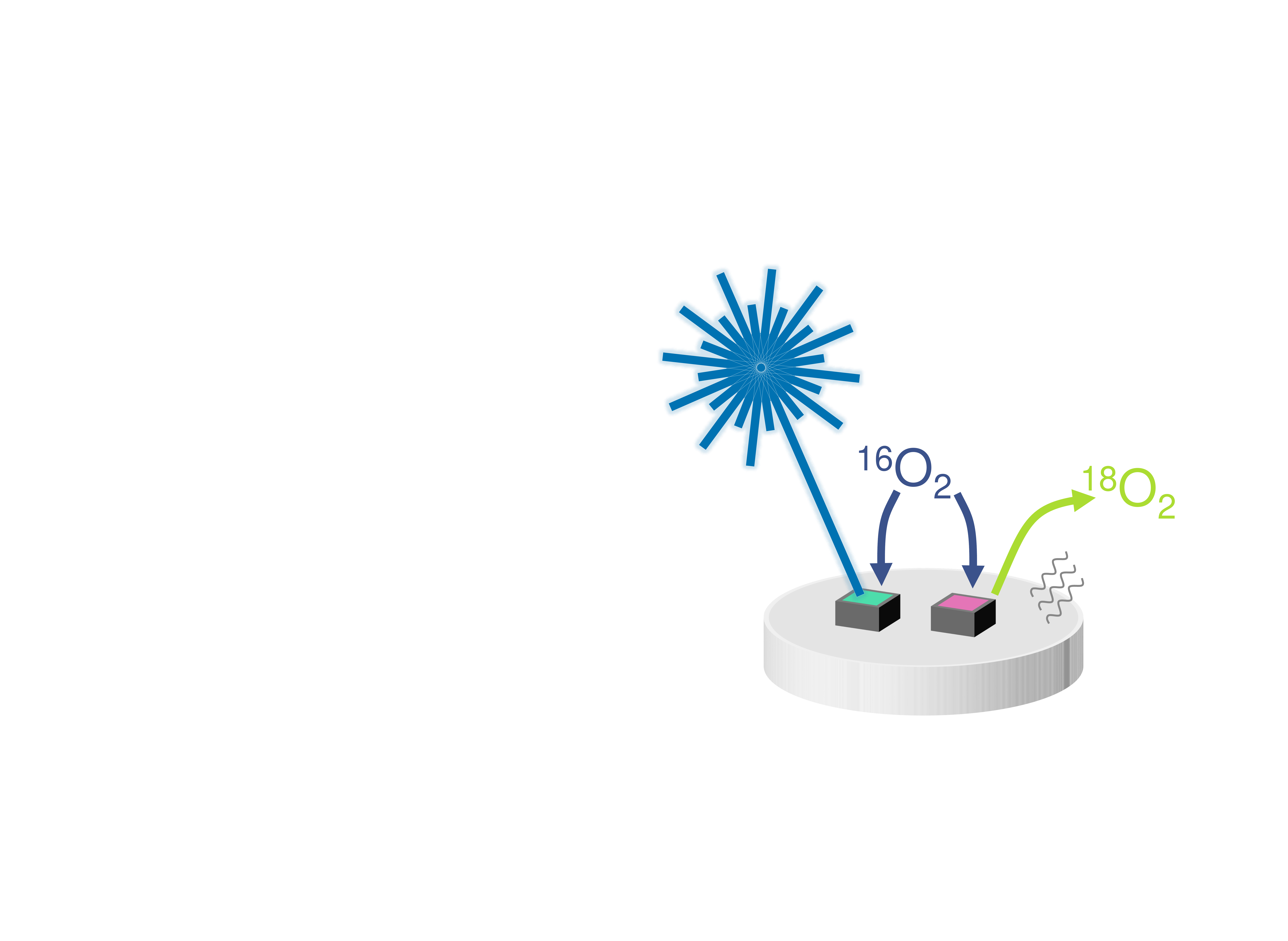}
			\put(0,0){\includegraphics[width=\textwidth, unit=1mm]{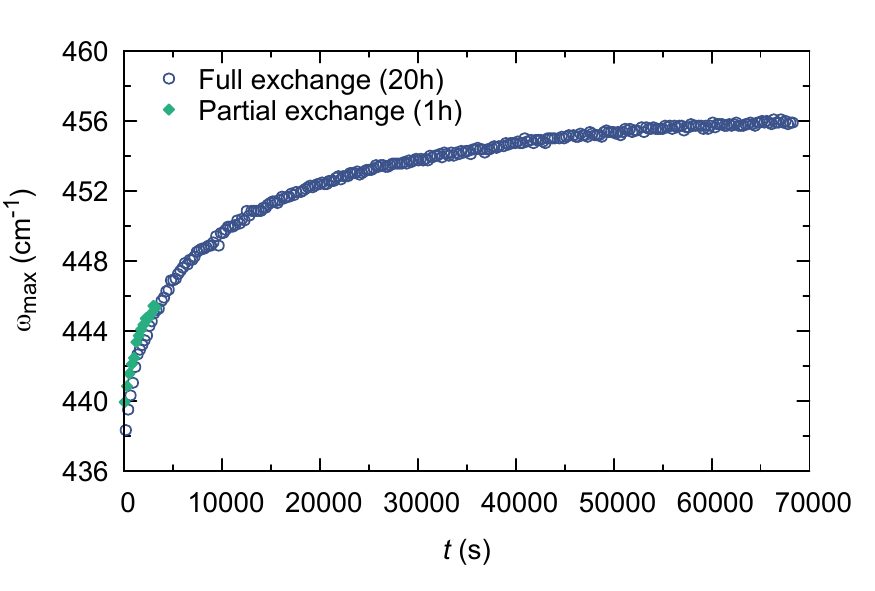}}
			\put(-2.5,55){(a)}
		\end{overpic}
	\end{minipage}
	\begin{minipage}[b]{0.49\textwidth}
		\begin{overpic}[width=\textwidth, unit=1mm]
			{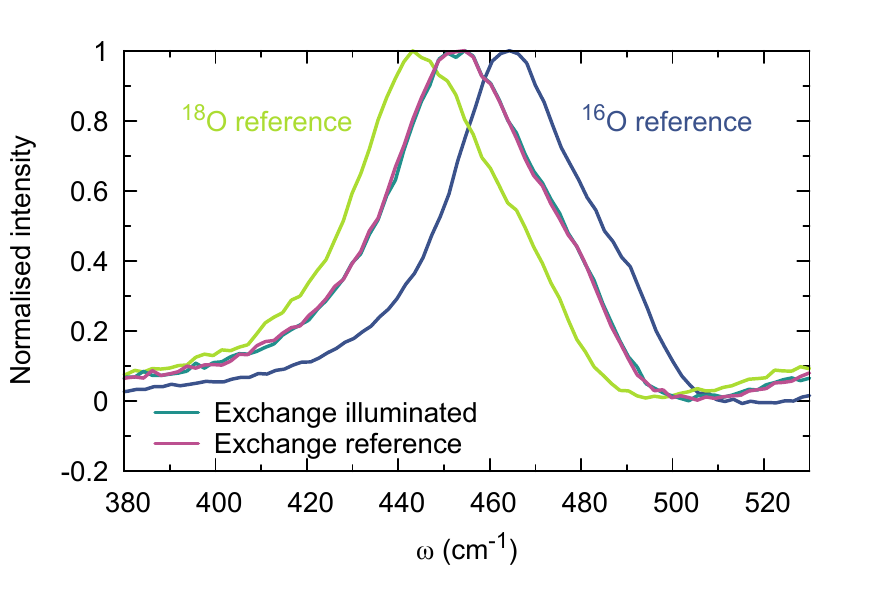}
			\put(-2.5,55){(b)}
		\end{overpic}
	\end{minipage}
	\caption[]{Evaluation of the potential influence of laser on exchange kinetics via direct laser heating or photoenhanced activity.
	The effect of laser light irradiation either due to direct laser heating or photoenhanced activity was evaluated by back-exchanging two samples simultaneously at 400\,°C for 3600\,s, with only one sample being exposed to focused laser light, as schematically shown in (a). The short annealing time leads to a partially back-exchanged state, which allows to quantify differences between the illuminated and reference samples after the back-exchange. Figure (a) compares the \textit{in situ} transients of a partial and a full back-exchange, showing the reproducibility of the measurement. The room temperature shifts of the two partially exchanged samples are depicted in (b) and compared to their initial $^{18}$O state and a fully back-exchanged $^{16}$O sample. Both samples exhibit the same shift, and therefore exchange kinetics of CGO are not notably enhanced under illumination and laser induced heating is negligible under chosen measurement conditions.}
	\label{fig:SI:Raman_CGO_photoenhanced}
\end{figure*}

The strengths of IERS for the analysis of surface limited kinetics are demonstrated by the study of CGO thin films. The \textit{in situ} isotopic shift of the Raman spectrum is shown in Fig.~\ref{fig:insitu_Raman}(a) for a back-exchange experiment in dry air atmosphere at 400\,°C. The shift of the T$_\text{2g}$ mode to higher wavenumbers is readily visible.
As the FWHM of the T$_\text{2g}$ band varies less than $5$\,\% over the course of the transition, see Fig.~\ref{fig:SI:devation_fit}(e), its influence on the determination of the peak position can be neglected. It is noteworthy that the employed fast heating ramp of 100\,°C\,min$^{-1}$ did not impact the integrity of the CGO thin films, as shown by SEM imaging after a thermal cycle, see Fig.~\ref{fig:SI:SEM_no_crack}. 
Also, no laser related activation or heating were found in the studied CGO samples, as shown in Fig.~\ref{fig:SI:Raman_CGO_photoenhanced}.


The gradual hardening of the normalized T$_\text{2g}$ mode is shown in Fig.~\ref{fig:insitu_Raman}(b) for different temperatures. Normalization is performed with respect to the theoretical mode position before any back-exchange, $\omega(t=0)$. A clear thermal activation of the exchange process is observed. 
The ($\omega_\text{max}$-$T$) diagram in Fig.~\ref{fig:insitu_Raman}(c) shows the transitions from the initial $^{18}$O to the final $^{16}$O state, whereas the expected temperature evolution of $\omega(t=0)$ and $\omega(t=\infty)$ is indicated with dashed lines. 
At high temperatures ($T\ge450$\,°C), the initial part of the transients cannot be resolved, as the time scale of the isotopic exchange is comparable to the acquisition time.

\begin{figure*}[t]
	\centering
	\begin{overpic}[width=175mm, unit=1mm]
		{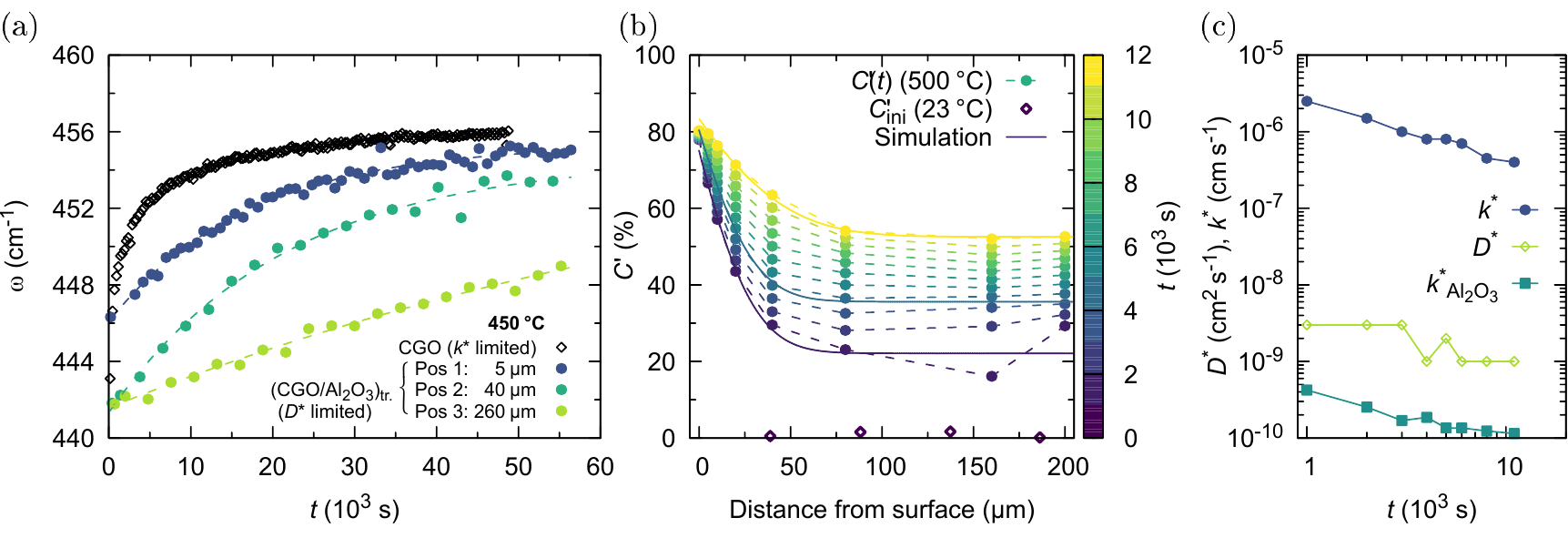}
	\end{overpic}
	\caption[]{\textbf{IERS for diffusion limited transport process}: In-plane diffusion measurements on thin film CGO sample with a 40\,nm Al$_2$O$_3$ capping layer:
		(a) \textit{In situ} Raman mode shift measured at 450\,°C of CGO/Al$_2$O$_3$ with a trench (coloured circles), measured at different in-plane distances from the open surface and uncoated CGO (open diamonds) (b) Normalised isotopic fraction as function of in-plane distance from open surface at 500\,°C. Dashed lines connect interpolated data points at discrete times, $t_i$. Solid lines correspond to FEM simulations (COMSOL) for $t_i=1000$, 4000 and 11000\,s, using three parameters ($k^*$, $D^*$ and $k^*_{\text{Al}_2\text{O}_3}$).
		(c) Transport parameters as function of time, obtained by FEM simulation of \textit{in situ} data at 500\,°C}
	\label{fig:inplane_diffusion}
\end{figure*}

The experimental data is modelled using Eq.~\ref{eqn:res:Fick} with a single and two parallel exponentials, as well as Eq.~\ref{eqn:res:k_expansion_quadr} derived for non-linear reaction rates, as shown for the back-exchange transient at 400\,°C in Fig.~\ref{fig:insitu_Raman}(b) with a dotted ($\tau_\text{single}$), dashed ($\tau_\text{parallel}$) and green dashed-dotted ($\tau_\text{non-lin.}$) line, respectively. 
The simplest case of a single exponential deviates considerably from the \textit{in situ} data, whereas the other two models adequately fit the experimental transients. The deviations of the different models from the experimental data are shown in Fig.~\ref{fig:SI:devation_fit}. 
The surface exchange coefficients of the $\tau_\text{parallel}$ model are shown in Fig.~\ref{fig:insitu_Raman}(d) (open triangles). Both, $k^*_\alpha$ and $k^*_\beta$, follow an Arrhenius type behaviour with an activation energy, $E_\text{A}$, of $0.9\pm0.1$\,eV.
This $E_\text{A}$ is comparable to values reported in literature for Gd doped CeO$_2$ within the analysed temperature range \cite{Schaube2019,Kowalski2009}. Due to fast kinetics at 500\,°C, only a single surface exchange coefficient could be estimated, which approximately falls onto the extrapolated Arrhenius law of $k^*_\alpha$.

For samples not fully back-exchanged (\textit{i.e.} ${^{18}C}$ > 10\,\%), it is possible to estimate a $k^*$ value based on \textit{ex situ} measurements of the ${^{18}C}$ after the thermal annealing by analysing the room temperature Raman shift and measuring the ${^{18}C}$ by ToF-SIMS. This conventional approach can estimate only a single time constant, which is shown as well in Fig.~\ref{fig:insitu_Raman}(d). 
Interestingly, these values fall onto the lines of $k_{\alpha}^*$ and $k_{\beta}^*$ depending on their annealing time. For short annealing durations of less than one hour, fast $k^*$ values are found (open diamonds and open circles), while for an annealing of 20\,h at 400\,°C a slow $k^*$ value is obtained (closed symbols, compare as well with SIMS depth profiles in Fig.~\ref{fig:Analysis_iso_shift}(a)). 
A similar effect was observed for the initial $^{18}$O exchange in CGO.
We can therefore conclude that simple linear kinetics indeed fail to model the observed processes and the time resolution of \textit{in situ} IERS provides a richer insight into the physical chemical processes than conventional approaches. 

Fig.~\ref{fig:CGO_k_literature} summarises surface exchange coefficients from this work and from the literature on different Gd doped ceria samples, ranging from 10 to 31\,mol\%. Activation energies range from 2.5 -- 4\,eV at high temperatures, to around 1.5\,eV at intermediate temperatures and down to 0.5 -- 1\,eV at low temperatures, while $k^*$ varies between two to three orders of magnitude for different sample types and analysis techniques, including IEDP and pulsed isotope exchange (PIE). 
Note that presented literature results are all obtained for $^{18}$O exchange experiments, while in this work we have studied \textit{in situ} isotope back-exchanges to minimise the consumption of expensive $^{18}$O tracer gas. However, tracer exchange and diffusion experiments are expected to be equivalent in both directions \cite{Cooper2017} and given a suitable temperature cell, IERS can readily be applied as well for the \textit{in situ} analysis of the $^{18}$O exchange.

Potential interactions with water vapour \cite{Waldow2020}, $p$O$_2$ dependence of $k^*$ \cite{Lane2000} and the presence of catalytically active contaminations \cite{Schaube2019} have been discussed previously in the literature, but could not entangle the observed scatter of surface exchange coefficients, which is commonly found for many different fluorite and perovskite-type materials. It is interesting to note that $k^*$ values obtained by Yoo via PIE connect well to our fast $k^*_\alpha$ values, where exchange times are very short (10\,ms -- 10\,s), while the exchange experiments of Kowalski \cite{Kowalski2009} with longer annealing times (4\,h) compare well with the slower \textit{in situ} time constants (i.e. $k^*_\beta$). 

In a scenario of a partially blocking surface (\textit{e.g.} inert particles on the surface of the CGO film), one of the two apparent \textit{in situ} $k^*$ values could correspond in reality to an in-plane diffusion process \cite{Preis2019}, with diffusion coefficients in the range of $1\cdot10^{-14}$ -- $1\cdot10^{-17}$\,cm$^2$s$^{-1}$. These values are orders of magnitude lower than $D^*$ values obtained for CGO by \textit{in situ} Raman spectroscopy or reported in literature. Thus it is implausible that one of the two parallel processes relates to a diffusion mechanism.

\begin{figure*}[b]
	\centering
	\begin{minipage}[b]{0.49\textwidth}
		\begin{overpic}[width=\textwidth, unit=1mm]
			{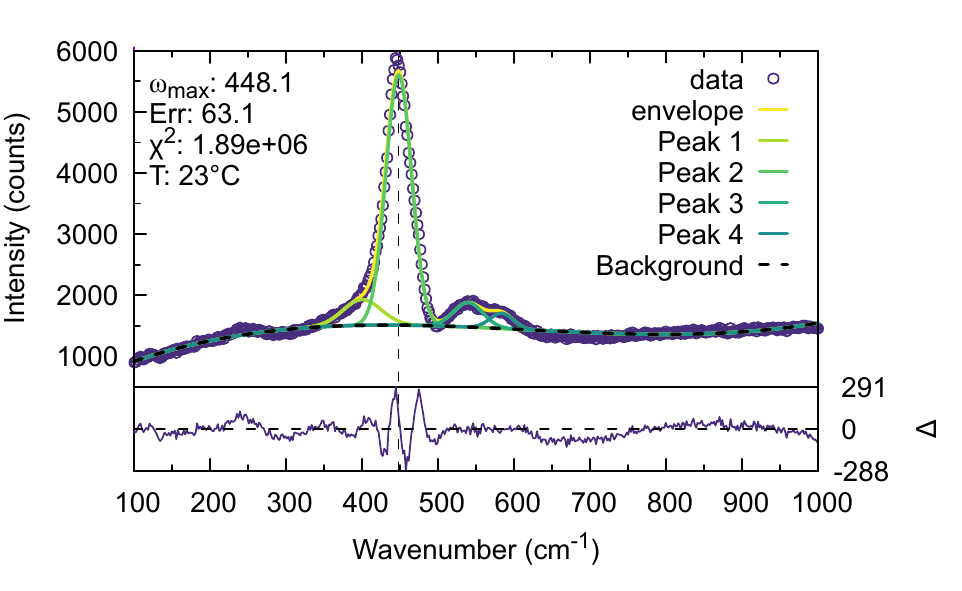}
			\put(-2.5,51){(a)}
		\end{overpic}
	\end{minipage}
	\begin{minipage}[b]{0.49\textwidth}
		\begin{overpic}[width=\textwidth, unit=1mm]
			{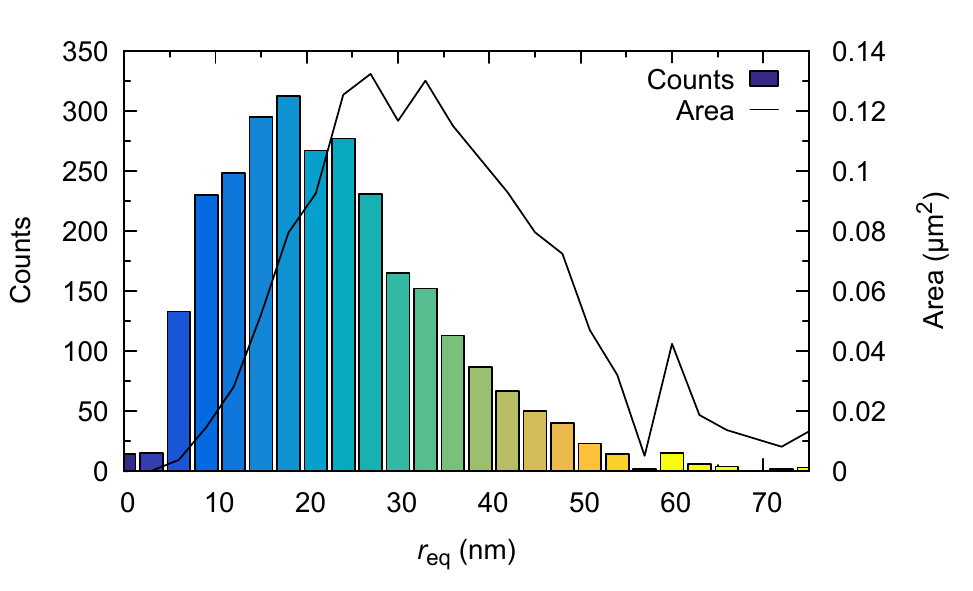}
			\put(-2.5,51){(b)}
		\end{overpic}
	\end{minipage}
	\caption[]{(a) Raman spectrum of $^{18}$O enriched CGO thin film at room temperature. It can be modelled using four Voigt profiles. In particular, we arbitrarily used two lines for the fit of the T$_\text{2g}$ mode, in order to take into account its asymmetry. This way of proceeding effectively allowed an accurate determination of the frequency of this mode ($\omega_\text{max}$, marked with a dashed line), which was the main information sought. Note that the difference between the position of the maximum of the envelope and the main mode is typically less than 1\,cm$^{-1}$, and hence within the error of the measurement itself. The background is corrected by fitting a 3\textsuperscript{rd} order polynomial curve to the data within the 100 - 300 and 700 - 1000\,cm$^{-1}$ $\omega$ ranges. The lower panel shows the difference between the measured intensity and the envelope. (b) Grain size distribution of polycrystalline CGO thin film on Pt/Si and the corresponding area per grain size. $r_\text{eq}$ corresponds to the radius of an equivalent circle with the same area as the grain.}
	\label{fig:SI_fitting}
\end{figure*}

On the other hand, Rutman \textit{et al.} found a dependence of $k^*$ on the grain size of Gd doped CeO$_2$ nano powder \cite{Rutman2014}, increasing by a factor of five from 5 to 50\,nm. As shown in Fig.~\ref{fig:SI_fitting}(b), the samples studied here show a broad grain size distribution, which could cause the deviation from a single exponential behaviour. This dependence could be potentially addressed by novel techniques based on isotope exchange coupled to atomic probe tomography with nanometric resolution \cite{Baiutti2021}, which is beyond the scope of this work.

\subsection*{IERS for diffusion limited transport processes}
\begin{figure}[t]
	\centering
	\begin{overpic}
		{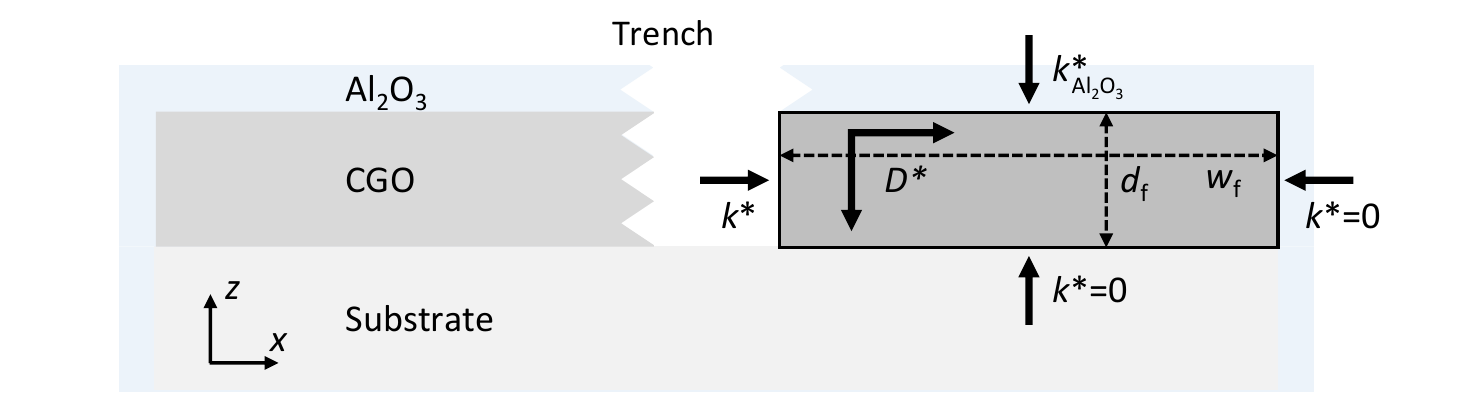}
	\end{overpic}
	\caption[]{Cross section of sample architecture for diffusion limited IERS and 2D geometry used in FEM simulations (framed dark grey area). The COMSOL model employs two different $k^*$ parameters for the top and side surface of the trench and isotropic oxygen diffusion ($D^*$). $k^*$ is defined via a linear reaction rate: $J^* = -D^* \frac{\partial {^{18}C}(x,t)}{\partial x}\Bigr\rvert_\text{surface}=k^*({^{18}C}_\text{gas}-{^{18}C}_\text{surface}(t))$. Oxygen exchange is inhibited from the bottom and the right side. $d_\text{f}$ and $w_\text{f}$ are film thickness and half width of the sample.}
	\label{fig:SI_COMSOL}
\end{figure}

The capability of IERS for the study of diffusion limited reactions in an unprecedented manner, is demonstrated using a CGO/Al$_2$O$_3$ model system, where oxygen exchange is enabled only at a trench along the centerline (see schematic in Fig.~\ref{fig:SI_COMSOL}).
Time resolved isotope diffusion profiles are obtained via consecutive line scans perpendicular to the trench, as shown in Fig.~\ref{fig:inplane_diffusion}(a) at 450\,°C.
At all $x$ positions, a smooth transition can be observed, while the Raman mode shift becomes much slower with increasing distance to the open surface, due to the increasing diffusion length.
In comparison to an un-capped CGO sample (open diamonds, surface limited process) kinetic differences are evident.
In a next step, each transient was interpolated via curve fitting, as indicated with dashed lines, to obtain $\omega(x_i,t_i)$ at discrete times, $t_i$ for all measured positions $x_i$, which can be converted into ${C'}$ using the VCA approximation.
The resulting isotopic diffusion profiles are shown in Fig.~\ref{fig:inplane_diffusion}(b). 
Exchange and diffusion coefficients ($k^*$ and $D^*$) were retrieved by comparing experimental data to FEM simulations (COMSOL, the used 2D geometry is shown in Fig.~\ref{fig:SI_COMSOL}).
Notably, ${C'}$ is flat for $x>80$\,µm, but increases with time. In the simplest case this can be understood as a small leakage of oxygen through the alumina capping and can be considered by the introduction of an additional exchange coefficient, $k^*_{\text{Al}_2\text{O}_3}$, at the top surface.
Representative least square solutions are shown for $t_i=$1000, 4000 and 11000\,s with solid lines in Fig.~\ref{fig:inplane_diffusion}(b). Resulting diffusion coefficients ($D^*\approx2\cdot10^{-9}$\,cm\,s$^{-2}$) match well with literature values extrapolated to 500\,°C ($D^*_\text{lit.}=1\cdot10^{-9}$\,--\,$1\cdot10^{-8}$\,cm\,s$^{-2}$) \cite{Manning1996,Ruiz-Trejo1998} and $k^*\approx 4\cdot10^{-7}$\,cm\,s$^{-1}$ is in reasonable agreement with values reported above. The exchange through the capping layer is orders of magnitude slower, \textit{i.e.} $k^*_{\text{Al}_2\text{O}_3}\approx 2\cdot10^{-10}$\,cm\,s$^{-1}$.

Interestingly, diffusion profiles at different times cannot be fitted with the same set of transport parameters ($k^*$, $k^*_{\text{Al}_2\text{O}_3}$ and $D^*$) but must be optimized for each time. Evolution with time of the best matching values is shown in Fig.~\ref{fig:inplane_diffusion}(c). Similar to the surface limited case described above, we find non-constant diffusion and exchange coefficients, which may again point towards non-linear thermodynamics already in the simple case of a purely entropic driving force in tracer exchange experiments.
Possessing the capability to address this type of questions renders one of the main advantages of IERS.



\begin{figure*}[t]
	\centering
	\begin{overpic}[width=175mm, unit=1mm]
		{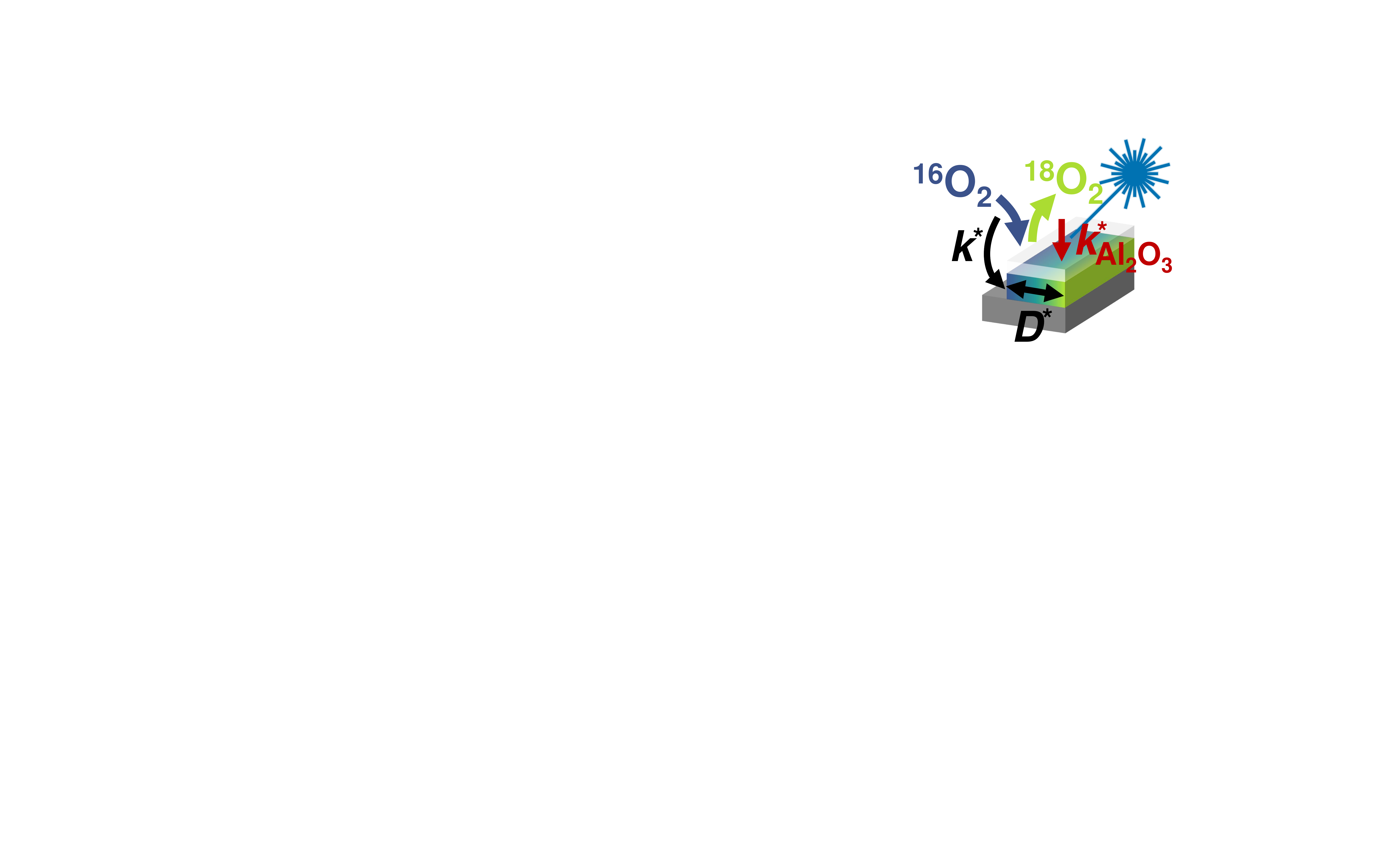}
		\put(0,0){\includegraphics[width=175mm, unit=1mm]{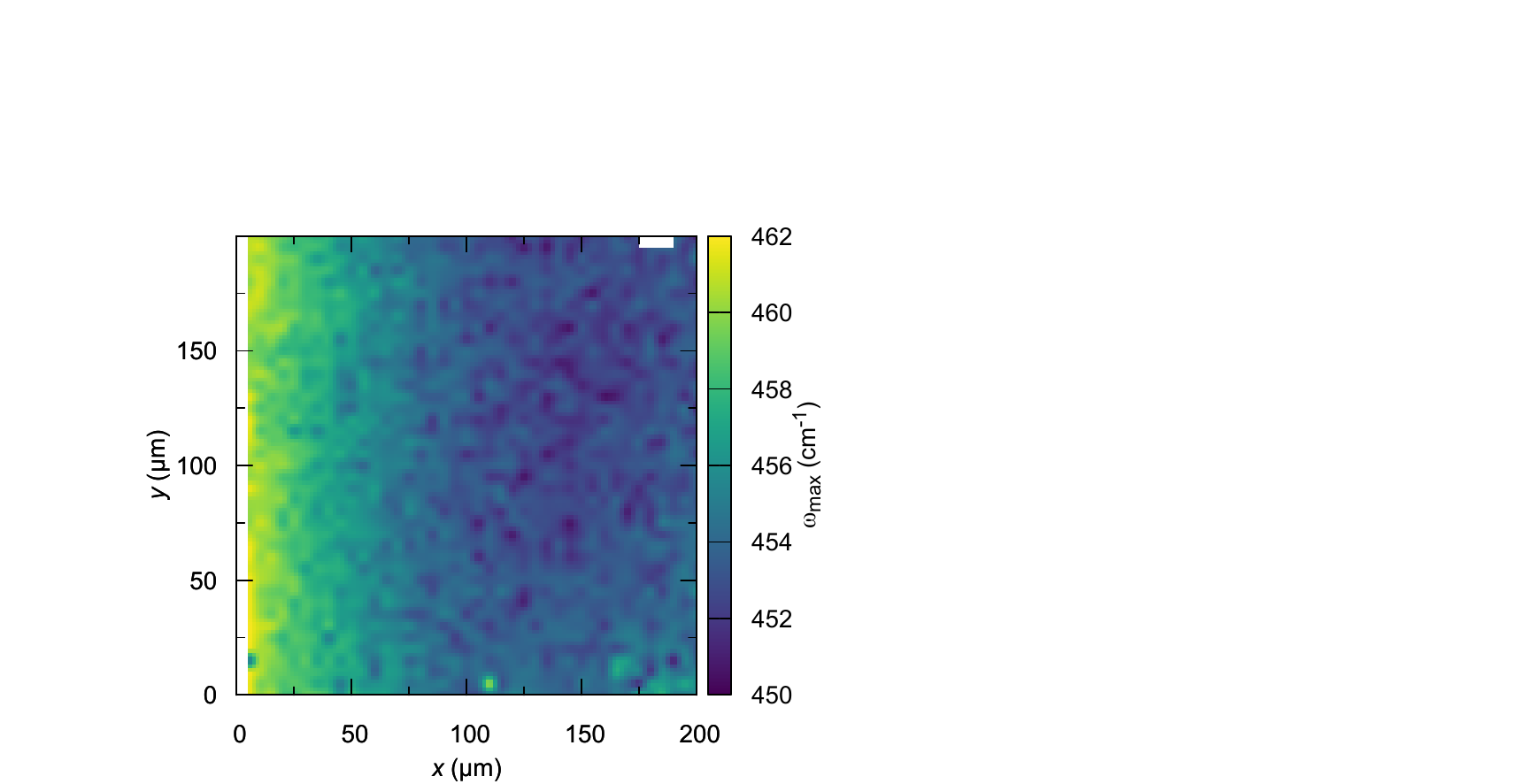}}
		\put(0,0){\includegraphics[width=175mm, unit=1mm]{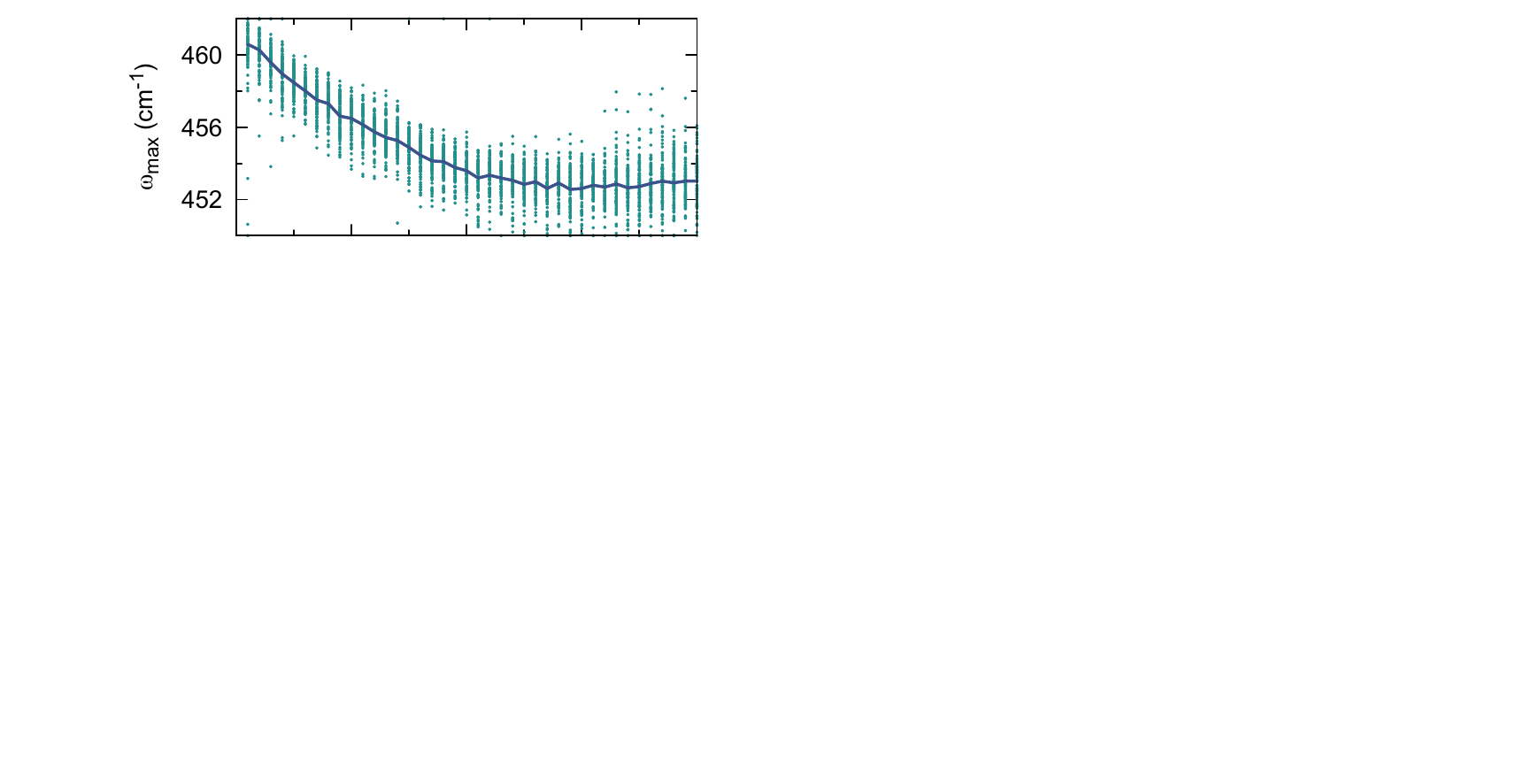}}
		\put(0,0){\includegraphics[width=175mm, unit=1mm]{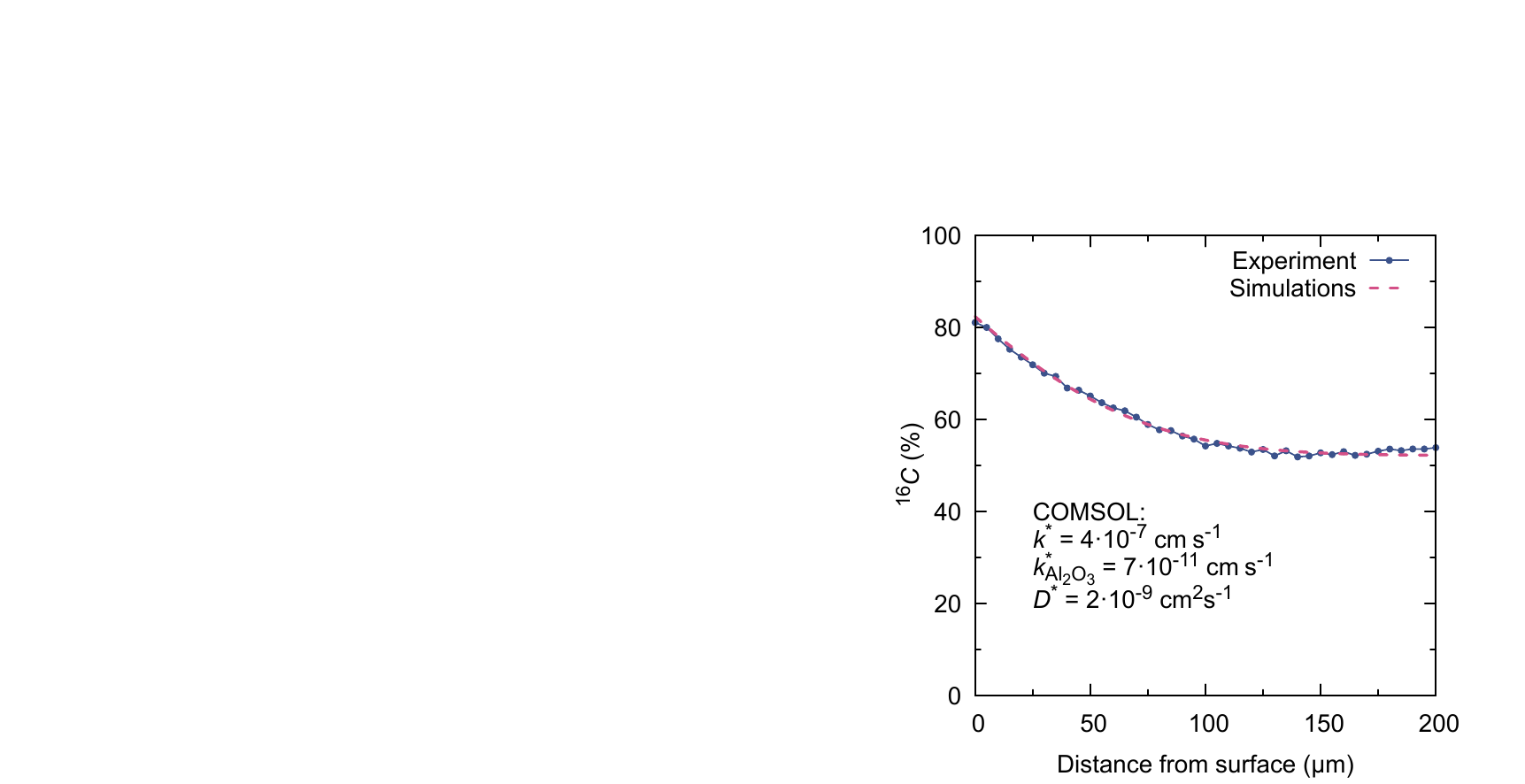}}
		\put(15,87){(a)}
		\put(100,87){(b)}
	\end{overpic}
	\caption[]{
	\textbf{\textit{Ex situ} analysis of in-plane isotopic diffusion profile:} Trenched CGO/Al$_2$O$_3$ film back-exchanged at 500\,°C for 12300\,s. (a) Surface mapping using Raman spectroscopy. The solid line in the upper panel corresponds to the average of all line scans. (b) ${^{16}C}$ data obtained from the average mode shift and via FEM simulations (COMSOL) using a surface exchange and diffusion coefficient, $k^*$ and $D^*$, for CGO and an additional surface exchange coefficient, $k^*_{\text{Al}_2\text{O}_3}$, to account for oxygen leakage through the capping layer, as schematically shown in the sketch on top.}
	\label{fig:SI:inplane_diffusion}
\end{figure*}

Finally, \textit{in situ} measurements are validated by room temperature Raman mapping of the surface after the back-exchange, as shown in Fig.~\ref{fig:SI:inplane_diffusion}(a). 
The isotopic in-plane gradient is readily visible via the wavenumber shift. The average line profile (solid line in the upper panel in Fig.~\ref{fig:SI:inplane_diffusion}(a)) is used to obtain transport properties via FEM simulations. The obtained values ($k^*=4\cdot10^{-7}$\,cm\,s$^{-1}$, $k^*_{\text{Al}_2\text{O}_3}=7\cdot10^{-11}$\,cm\,s$^{-1}$ and $D^*=2\cdot10^{-9}$\,cm\,s$^{-2}$) are in perfect agreement with the figures from \textit{in situ} measurements, obtained for the last diffusion profile ($t_i$=11000\,s). This validates that isotopic exchange Raman spectroscopy is a powerful tool for the characterisation of mass transport properties of oxide materials under \textit{in situ} conditions.

\section*{Conclusions}
Novel isotope exchange Raman spectroscopy (IERS) allows to directly measure tracer surface exchange and self-diffusion coefficients \textit{in situ} with time and spatial resolution, not accessible by means of conventional techniques.
We have used IERS to measure transport properties and corresponding activation energies for Gd doped CeO$_2$ thin films. 
The obtained values are in good agreement with classical \textit{ex situ} approaches and previously reported values. However, thanks to the time resolution of IERS, we were able to gain additional information depth opening up questions on the validity of simple linear irreversible reaction kinetics, which can be further addressed using this new methodology. 

The strengths of IERS are its simple setup, fast and easy data acquisition and analysis, economic benefits and resource efficiency, and therefore its wide applicability for the study of functional oxides for energy applications and beyond as a function of various process parameters under \textit{in situ} and \textit{operando} conditions.

\section*{Experimental}
Ce$_{1-x}$Gd$_{x}$O$_{2}$ ($x=0.2$) thin films were fabricated by large-area pulsed laser deposition (PVD Systems, PLD 5000) equipped with a 248\,nm KrF excimer laser (Lambda Physics, COMPex PRO 205) under the following conditions: temperature 700\,°C, oxygen pressure $7\cdot10^{-3}$\,mbar, target to substrate distance 90\,mm, laser fluency $\approx$1.2\,Jcm$^{-2}$, 10\,Hz repetition rate.
The oxide thin films were deposited on top of 10$\times$10\,mm$^2$ Si (111) wafers, coated with a Pt (150\,nm)/TiO$_2$ (40\,nm)/SiO$_2$ (500\,nm) multilayer (LETI, Grenoble, France). 
After deposition, samples were cut into several pieces using a water cooled diamond saw. 
Thin (40-90\,nm), conformal Al$_2$O$_3$ capping layers were deposited by atomic layer deposition (ALD) at 180\,°C on top of the $^{18}$O filled CGO thin films. Trenches were opened across the Al$_2$O$_3$ coating and CGO layer at the center of the sample with a diamond tip, exposing the out-of-plane CGO surface to the atmosphere.

For the isotope exchange, samples were placed in a sealed quartz tube, evacuated and purged with Ar for three times, before $^{18}$O enriched oxygen was introduced ($c(^{18}O)=98$\,\%, $p=0.2$\,atm, CK Isotopes, UK).
Samples were rapidly heated by rolling on a tubular furnace with an approximate ramp of 100\,°C\,min$^{-1}$ to 640\,°C. After 0.5\,h, the furnace was rolled off. Pre-equilibration steps in natural oxygen, as commonly performed for tracer diffusion experiments, were omitted, as a high $^{18}$O isotopic fraction was desired. For CGO, the oxygen vacancy concentration within the studied temperature window is fixed by the amount of Gd doping \cite{Yashiro2002,Wang1998}. A temperature difference between the initial exchange and the subsequent back-exchange, is therefore not expected to cause an additional chemical driving force.

The chemical and isotopic composition was obtained by SIMS analysis, which was performed using a ToF-SIMS V instrument (ION-TOF GmbH Germany) equipped with a Bi liquid metal ion gun (LMIG) for analysis and a caesium (Cs$^+$) gun for sputtering. Negative secondary ions were collected and data was obtained in burst mode operation (6 pulses), applying Poisson correction. 
Charge effects were compensated by means of a 20\,eV pulsed electron flood gun. 
Depth profiling was performed by alternating sputtering of a 300$\times$300\,mm$^2$ surface area with the Cs$^+$ ion beam (2\,keV, 110\,nA) and chemical analysis with the Bi$^{3+}$ primary ion beam (25\,keV, 0.25\,pA, rastered surface area: 50$\times$50\,µm$^2$). Generally two SIMS profiles per sample were recorded to confirm its homogeneity.

Raman spectroscopy measurements were performed using a Jobin Yvon/Horiba Labram spectrometer equipped with a liquid nitrogen cooled CCD detector (Jobin Yvon-Horiba Spectrum One CCD3000V). The blue excitation wavelength (488\,nm) of an Ar$^+$ laser was focused onto the surface with an approximate spot size of 1\,µm$^2$. A 50$\times$ (long working distance) and a 100$\times$ objective were used for \textit{in situ} studies and room temperature measurements, respectively.
The maximum laser power on the sample surface did not exceed 0.7-0.8\,mW. Spectra were calibrated at room temperature using a silicon reference sample with a theoretical position of 520.7\,cm$^{-1}$.
The background was subtracted for each spectrum using a 3\textsuperscript{rd} order polynomial.

For \textit{in situ} back-exchange measurements, a temperature cell (Linkam THMS 600 and Nextron MPS-CHH) was mounted onto the Raman stage. The cells are equipped with a ceramic heater (diameter of 1" Linkam, 0.5" Nextron) and a sapphire window for Raman measurements. For spatial resolution, the temperature cell was mounted onto a motorized linear stage.
The temperature cell was preheated for several hours before loading a sample to thermally equilibrate the setup and avoid the loss of focus due to thermal expansion during the measurement.
Fast heating ramps of up to 100\,°C\,min$^{-1}$ were used to minimise any isotopic exchange during the heating.
Automatic Raman acquisition was typically started within 30-60\,s after reaching the exchange temperature, with acquisition times of 120-240\,s.
Isotopic back-exchanges were performed under flowing dry air ($<100$\,ml/min) of natural $^{18}$O composition at atmospheric pressure.

\section*{Data availability}
All sample datasets and materials related to this work are made available under CC BY 4.0 license in the zenodo repository: 10.5281/zenodo.7072918.
\section*{Acknowledgements}
This work has received funding from the European Union's Horizon 2020 research and innovation program under grant agreement no. 824072 (Harvestore) and under the Marie Skłodowska-Curie grant agreements no. 840787 (Thin-CATALYzER) (for F.B) and 
no. 746648  (PerovSiC) (for D.P.). We acknowledge Stefano Ambrosio for his help with initial IERS measurements and Florence Robaut and Laetitia Rapenne for FIB lamella preparation and TEM analysis. 
This research has benefited from characterization equipment of the Grenoble INP - CMTC platform supported by the Centre of Excellence of Multifunctional Architectured Materials "CEMAM" n°ANR-10-LABX-44-01 funded by the "Investments for the Future" Program. In addition, this work has been performed with the help of the “Plateforme Technologique Amont” de Grenoble, with the financial support of the “Nanosciences aux limites de la Nanoélectronique” Fundation" and CNRS Renatech network. Chevreul institute (FR 2638), the French ministry of research, the Région Hauts de France and FEDER are acknowledged for supporting and funding the surface analysis platform (ToF-SIMS). 

\section*{Author contributions statement}
Conceptualization: MB, AS; sample preparation: FB, FC, AS; investigation: AS, CP, MM; formal analysis: AS; methodology: AS, MB, OC, DP, CJ, MM; original draft: AS; review \& editing: all authors 

\bibliography{library}

\begin{thebibliography}{10}
\bibitem{footnote1}
\bibinfo{title}{The opened side surface must be sufficiently large to avoid any
  out-of-plane isotopic gradient, especially relevant for bulk materials with
  anisotropic diffusion properties. alternatively to cutting a trench, one side
  surface can be left uncapped in bulk samples.}
\bibitem{footnote2}
\bibinfo{title}{The net flux vanishes at equilibrium, thus $k^*_0$ must be 0.}
\end{thebibliography}

\beginsupplement
\clearpage
\onecolumn
\section*{Supplementary information}
\begin{Large}{Isotope Exchange Raman Spectroscopy (IERS): a novel technique to probe physicochemical processes \textit{in situ}}\end{Large}
\\
\\
Alexander Stangl$^{1,*}$, Dolors Pla$^1$, Caroline Pirovano$^2$, Odette Chaix-Pluchery$^1$, Federico Baiutti$^{3,4}$, Francesco Chiabrera$^3$, Albert Tarancón$^{3,5}$, Carmen Jiménez$^1$, Michel Mermoux$^5$, Mónica Burriel$^1$
\\
\\
$^1$Univ. Grenoble Alpes, CNRS, Grenoble-INP, LMGP, 38000 Grenoble France \\
$^2$Univ. Lille, CNRS, Centrale Lille, Univ. Artois, UMR 8181 – UCCS – Unité de Catalyse et Chimie du Solide, F-59000 Lille, France\\
$^3$Catalonia Institute for Energy Research (IREC), Barcelona, Spain\\
$^4$Departement of Materials Chemistry, National Institute of Chemistry, Hajdrihova 19, Ljubljana SI-1000, Slovenia\\
$^5$ICREA, 23 Passeig Lluis Companys, 08010 Barcelona, Spain\\
$^6$Univ. Grenoble Alpes, Univ. Savoie Mont Blanc, CNRS, Grenoble INP, LEPMI, 38000, Grenoble, France\\
$^*$alexander.stangl@grenoble-inp.fr, monica.burriel@grenoble-inp.fr

\end{document}